\documentclass{aa}  % print
\usepackage[T1]{fontenc}
\pdfoutput=1 
\usepackage{ae,aecompl}
\usepackage{amssymb,amsmath,amsfonts}
\usepackage{graphicx}
\graphicspath{{./figures/}}
\usepackage[]{hyperref}
\hypersetup{colorlinks=true,linkcolor=blue,citecolor=blue,filecolor=blue,
urlcolor=blue}

\def\PyBDSF/{{\sc PyBDSF}}

\begin{document} 

%%%%%%%%%%%%%%%%%%%%%%%%%%%%%%%%%%%%%%%%%%%%%%%%
\title{The LOFAR Two-metre Sky Survey\thanks{LoTSS}} 
\subtitle{III. First Data Release: Optical/infrared identifications and 
value-added catalogue} 

% Group 1
\authorrunning{Williams et~al.}
\titlerunning{LoTSS-DR1 optical identifications}
\author{W.~L.~Williams$^{1}$\thanks{E-mail: w.williams5@herts.ac.uk}, 
M.~J.~Hardcastle$^{1}$,
P.~N.~Best$^{2}$,
J.~Sabater$^{2}$,
J.~H.~Croston$^{3}$,
K.~J.~Duncan$^{4}$,
T.~W.~Shimwell$^{5,4}$,
H.~J.~A.~R\"{o}ttgering$^{4}$,
D.~Nisbet$^{2}$,
G.~G\"urkan$^{6}$,
%~group~2
L.~Alegre$^{2}$,
R.~K.~Cochrane$^{2}$,
A.~Goyal$^{7}$,
C.~L.~Hale$^{8}$,
N.~Jackson$^{9}$,
M.~Jamrozy$^{7}$,
R.~Kondapally$^{2}$,
M.~Kunert-Bajraszewska$^{10}$,
V.~H.~Mahatma$^{1}$,
B.~Mingo$^{3}$,
L.~K.~Morabito$^{8}$,
I.~Prandoni$^{11}$,
C.~Roskowinski$^{10}$,
A.~Shulevski$^{12}$,
D.~J.~B.~Smith$^{1}$,
C.~Tasse$^{13,14}$,
S.~Urquhart$^{3}$,
B.~Webster$^{3}$,
G.~J.~White$^{3,15}$,
% group 3
R.~J.~Beswick$^{9}$,
J.~R.~Callingham$^{5}$,
K.~T.~Chy\.zy$^{7}$,
F.~de~Gasperin$^{16}$,
J.~J.~Harwood$^{1}$,
M.~Hoeft$^{17}$,
M.~Iacobelli$^{5}$,
J.~P.~McKean$^{5,18}$,
A.~P.~Mechev$^{4}$,
G.~K.~Miley$^{4}$,
D.~J.~Schwarz$^{19}$,
R.~J.~van Weeren$^{4}$\\
(Affiliations can be found after the references)}
\institute{}
\date{Accepted November 9, 2018; received  June 5, 2018}

\abstract{
\noindent
The LOFAR Two-metre Sky Survey (LoTSS) is an ongoing sensitive, high-resolution 
120-168\,MHz survey of the northern sky with diverse and ambitious science 
goals. Many of the scientific objectives of LoTSS rely upon, or are enhanced by, 
the association or separation of the sometimes incorrectly catalogued radio 
components into distinct radio sources and the identification and 
characterisation of the optical counterparts to these sources. We present the 
source associations and optical and/or IR identifications for sources in the 
first data release, which are made using a combination of statistical techniques 
and visual association and identification. We document in detail the colour- and 
magnitude-dependent likelihood ratio method used for statistical identification 
as well as the Zooniverse project, called LOFAR Galaxy Zoo, used for visual 
classification. We describe the process used to select which of these two 
different methods is most appropriate for each LoTSS source.  The final 
LoTSS-DR1-IDs value-added catalogue presented contains 318,520 radio sources, of 
which 231,716 (73\%) have optical and/or IR identifications in Pan-STARRS and 
\textit{WISE}. The value-added catalogue is available on-line  at  
\url{https://lofar-surveys.org/}, as part of this  data release.
}

\maketitle

\defcitealias{Shimwell_2018}{DR1-I}
\defcitealias{Duncan_2018}{DR1-III}

\section{Introduction}
The true power of modern large radio surveys, which will reveal many millions of 
radio sources, lies in cross-matching them with surveys at different 
wavelengths, i.e. in identifying the multiwavelength counterparts of radio 
sources. This enables detailed statistical studies of the populations of 
extragalactic radio sources and their host galaxy properties. Over the last few 
decades, the cross-matching of large area radio surveys, in particular the 
National Radio Astronomy Observatory (NRAO) Very Large Array (VLA) Sky Survey 
\citep[NVSS;][]{Condon_1998} and the Faint Images of the Radio Sky at Twenty 
centimetres (FIRST) survey \citep{Becker_1995},  with large-scale optical 
spectroscopic surveys, such as the Sloan Digital Sky Survey 
\citep[SDSS;][]{York_2000,Stoughton_2002} and the 6 degree Field Galaxy Survey 
\citep[6dFGS;][]{Jones_2004}, have hugely improved our understanding of 
extragalactic radio sources. Matching these surveys has provided samples of many 
thousands of sources \citep[e.g.][]{Best_2005a,Mauch_2007}, which have allowed 
for detailed statistical studies of the radio source populations 
\citep[e.g.][]{Best_2005b,Best_2012,Janssen_2012}.

In the coming years, a number of wide area surveys will be carried out using the 
next generation of radio telescopes and telescope upgrades. These include  the 
LOw Frequency ARray (LOFAR;  \citealt{vanHaarlem_2013}) Two-metre Sky Survey 
(LoTSS; \citealt{Shimwell_2017}),  the VLA Sky Survey 
(VLASS\footnote{\url{https://science.nrao.edu/science/surveys/vlass}}), the  
Evolutionary Map of the Universe survey \citep[EMU;][]{Norris_2011} using  the 
Australian SKA Pathfinder \citep[ASKAP;][]{Johnston_2007}, and the WODAN survey 
\citep{Rottgering_2011} using the  APERture Tile In Focus \citep[APERTIF; 
][]{Verheijen_2008} upgrade on the Westerbork Synthesis Radio Telescope (WSRT). 
New large-area optical surveys are also in progress or planned. These include 
surveys with the Panoramic Survey Telescope and Rapid Response System 
\citep[Pan-STARRS;][]{Kaiser_2002,Kaiser_2010}, the Large Synoptic Survey 
Telescope \citep[LSST;][]{Ivezic_2008} and \textit{Euclid} 
\citep{Amendola_2016}. Deep X-ray surveys with eROSITA  are also planned 
\citep{Merloni_2012}. When combined, these next generation radio and 
multiwavelength surveys will provide samples orders of magnitude larger than 
currently available, reaching to substantially higher redshifts,  which will 
revolutionise our understanding of radio source populations through far more 
detailed statistical studies. 

Cross-matching surveys at different wavelengths is a well-established procedure 
in astronomy, albeit with some unresolved challenges. For many radio sources, 
including star-forming galaxies and some radio-loud active galactic nuclei 
(AGN), the radio emission is relatively compact and is coincident with the 
optical emission, allowing cross-matching through simple procedures, such as 
nearest neighbour (NN) matching or more complex automated statistical methods. 
However, problems of matching between the radio and optical are compounded by 
the complex nature of other radio sources, in particular spatially extended 
radio-loud AGN: these scientifically interesting complex-structured sources are 
very challenging to cross-match. 

A sensitive, high-resolution 120-168\,MHz survey of the northern 
sky, LoTSS, is already well under way. 
Using the High Band Antenna (HBA) system of LOFAR, the survey aims to reach a 
sensitivity of less than $0.1$~mJy~beam$^{-1}$ at an angular resolution of 
$\sim6\arcsec$ across the whole northern hemisphere.  The first data release 
(LoTSS-DR1), described in the accompanying paper \citep[][hereafter 
DR1-I]{Shimwell_2018}, covers 424 square degrees and includes over 300,000 
radio sources. While surveys like NVSS lack angular resolution and surveys like 
FIRST have problems with resolving out large-scale emission,  LoTSS is unique 
in retaining both high resolution and sensitivity to large-scale structures, 
which aids the process of cross-matching. Many of the scientific objectives of 
LoTSS rely upon, or are enhanced by, the identification and characterisation of 
the multiwavelength counterparts to the detected radio sources. In this paper 
we have made our first attempt at enriching our radio catalogues by identifying 
their optical/IR\footnote{In this paper we take optical/IR to mean the 
inclusive or, i.e. optical or IR or both.} counterparts, thereby enabling their 
photometric and spectroscopic redshifts to be determined. Accurate source 
redshifts allow physical properties such as luminosities and sizes to be 
determined, which in turn enables studies of the intrinsic properties of radio 
sources and their host galaxies\footnote{For examples of the broad range of 
science see the other papers in this special issue.}. Photometric redshift and 
rest-frame colour estimates for all the matched optical/IR sources are 
presented in the accompanying paper \citep[][hereafter DR1-III]{Duncan_2018}. 
Furthermore, future spectroscopic surveys such as WEAVE-LOFAR 
\citep{Smith_2016}, using the William Herschel Telescope Enhanced Area Velocity 
Explorer \cite[WEAVE;][]{Dalton_2012,Dalton_2014} multi-object and integral 
field spectrograph, will provide precise redshift estimates and robust source 
classification for large fractions of the LoTSS source population.

This paper is structured as follows. In Section \ref{sec:data} we give a brief 
summary of the LoTSS and optical/IR data used for the cross-matching. In Section 
\ref{sec:cataloging} we give an overview of the process of radio--optical 
cross-matching. The details of the statistical likelihood ratio (LR) technique 
are given in Section\ \ref{sec:cat:ml} and the full Zooniverse visual 
classification scheme is described in Section\ \ref{sec:cat:lgz}. In 
Section~\ref{sec:cat:flow} we present the decision tree that is used to decide 
which sources are identified by the likelihood ratio and visual classification 
methods. The final value-added catalogue is presented in Section\ 
\ref{sec:cat:final}, along with some of its basic properties. Finally, we 
summarise our work and discuss some possible future developments in Section\ 
\ref{sec:summary}.

Throughout this paper, all magnitudes are quoted in the AB system 
\citep{Oke_1983} unless otherwise stated.

\section{The radio and optical catalogues}
\label{sec:data}
\subsection{The LOFAR sample}
Details of the LoTSS first data release images and source extraction are given 
in \citetalias{Shimwell_2018} and we summarise the relevant points. The images 
cover 424 square degrees over\footnote{LoTSS-DR1 covers a region slightly larger 
than the HETDEX field, but with a few holes from four failed LOFAR pointings.} 
the Hobby-Eberly Telescope Dark Energy Experiment \citep[HETDEX;][]{Hill_2008} 
Spring Field (right ascension 10h45m to 15h30m and declination 
45$^\circ$00$\arcmin$ to 57$^\circ$00$\arcmin$). Direction-dependent calibration 
of the LOFAR data enabled imaging at the full resolution of 6$\arcsec$.  Source 
detection was performed on each mosaic image using the {\sc Python} Blob 
Detector and Source Finder  \citep[\PyBDSF/;][]{Mohan_2015}.  The background 
noise was estimated across the images using sliding box sizes of 30$\times$30 
synthesised beams, decreased to just 12$\times$12 synthesised beams near high 
signal-to-noise sources ($\geq$150) to more accurately capture the increase in 
noise over smaller spatial scales in these regions. Wavelet decomposition, with 
4 wavelet scales, was used to better characterise the complex extended emission 
present in the images. We set \PyBDSF/ to form islands with a 5$\sigma$ peak 
detection threshold and a 4$\sigma$ island threshold.
 Internally \PyBDSF/ fitted each island with one or more Gaussians that were 
grouped into discrete sources. The parameters we used for the source extraction 
(namely the box sizes for determining the background noise and the `group\_tol' 
parameter, for which we used a value of 10) were optimised through trial and 
error testing\footnote{This was done by visually examining the output catalogues 
overlaid on the LoTSS images prior to any of the visual classification presented 
in this paper.}. This allowed us to produce the best grouping of Gaussian 
components, i.e. to join up most compact double sources while not overproducing 
`blended' sources (incorrectly grouping separate sources as one source).  
Sources fitted with multiple Gaussians are identified in the  \PyBDSF/ source 
catalogue  by a value of `M' in the `S\_Code' column, those fitted by a single 
Gaussian have `S' in the `S\_Code' column, and a few tens of sources that are 
fitted by a single Gaussian, but lie within the same island as another source,  
have `C' in the `S\_Code' column. We treat `C' type sources the same as `M' type 
sources.

A final \PyBDSF/ source catalogue of the HETDEX region, containing 325,694 
entries, was produced, along with a final catalogue of all the Gaussian 
components of the \PyBDSF/  sources. In the following we refer to the 
source catalogue as the \PyBDSF/ source catalogue and the Gaussian 
component catalogue as the \PyBDSF/ Gaussian catalogue. Catalogue parameters 
refer to those from the \PyBDSF/ source catalogue, unless explicitly specified 
as the parameters from the \PyBDSF/ Gaussian component catalogue. 
\citetalias{Shimwell_2018} determined the positional accuracy of the catalogued 
sources to be within 0.2$\arcsec$.  

\subsection{The optical/infrared galaxy sample}
Deep and wide optical and IR data are available over the LoTSS-DR1 sky area 
from 
Pan-STARRS (in $grizy$ bands) and from the Wide-field Infrared Survey Explorer  
\citep[\textit{WISE};][]{Wright_2010}. The Pan-STARRS 3$\pi$ survey 
\citep{Chambers_2016}  covers the entire sky north of $\delta > -30^{\circ}$ 
with $5\sigma$ magnitude limits in the stacked $grizy$ images of $23.3, 23.2, 
23.1, 22.3$ and $21.4$\,mag, respectively. The typical point spread function 
(PSF) of the Pan-STARRS images is $\sim1-1.3${\arcsec}. The AllWISE catalogue 
\citep{Cutri_2013} includes photometry in the 3.4, 4.6, 12, and 22\,$\mu$m 
mid-infrared bands ($W1$, $W2$, $W3,$ and $W4$) for more than 747 million 
sources over the full sky. The $W1$ and $W2$ bands have significantly better 
sensitivity than the other two \textit{WISE} bands; the AllWISE catalogue 
completeness varies over the sky, but nominally it is $>95$\% complete for 
sources with $W1<19.8$, $W2<19.0$, $W3<16.67,$ and $W4<14.32$\,mag. The 
effective PSF for the \textit{WISE} images is  $6 - 6.5${\arcsec} in bands $W1$, 
$W2,$ and $W3$, and $\sim12${\arcsec} in $W4$.

We produced a combined Pan-STARRS--AllWISE catalogue over the LoTSS coverage 
area by matching sources in the two catalogues using the LR method, the details 
of which are given in Section \ref{sec:panstarrswise}. This combined catalogue 
includes sources with detections in only PanSTARRS or only AllWISE or both and 
is used for identifying the optical/near-infrared counterparts to LoTSS sources  
and in the determination of photometric redshifts and rest-frame colours 
\citepalias{Duncan_2018}.

For some large optical galaxies we make use of other earlier all-sky surveys, in 
particular, we use the SDSS DR-12 catalogue \citep{Alam_2015} and the Two Micron 
All Sky Survey \citep[2MASS;][]{Skrutskie_2006} extended source catalogue 
\citep[2MASX;][]{Jarrett_2000}. We refer only to source names in these 
catalogues. 

\section{Radio-optical cross-matching}
\label{sec:cataloging} 
Our objectives throughout this paper are essentially to correctly `associate' 
radio sources -- that is, to decide which sources found by the source finder 
belong together as components of one physical source  and which are separate 
sources that have been incorrectly associated by the source finder -- and to 
`identify' them -- that is, to find the best possible optical/IR counterpart 
where one exists.

The \PyBDSF/ catalogue is not a perfect representation of radio
sources. In addition to the unambiguous complete sources, this catalogue 
contains
a mixture of (i) blended sources, where distinct nearby sources have
been incorrectly associated as one source; (ii) separate components
of distinct sources, where a single source has been catalogued in
multiple entries because there is no contiguous emission between its
components (for example in the case of separate lobes of radio galaxies) so
that the true association is not recovered by the source finder; and (iii) 
spurious emission
or artefacts. We aim to produce a catalogue of real, correctly
associated radio sources and to provide their Pan-STARRS/\textit{WISE} 
counterparts, where possible. We handle the counterpart identification and 
possible association or separation of incorrectly catalogued components in two 
ways; we use a separate decision process to determine which of the two methods 
to use based on the properties of the radio sources. 

The first method determines  the presence or absence of a counterpart 
statistically. For this we use the LR, i.e. the ratio of the probability of a 
particular source being the true counterpart to that of it being a random 
interloper. This method is described in detail in Section \ref{sec:cat:ml}, and 
the specific application to this data set is described in Section\ 
\ref{sec:ml:lrapp}. Initially we determine the LR counterparts for all sources 
in the \PyBDSF/ catalogue with sizes smaller than $30${\arcsec} as well as for 
all the \PyBDSF/ Gaussian components smaller than $30${\arcsec}. These can be 
incorrectly combined into sources by \PyBDSF/ and individually have superior LR 
matches by themselves; for sources and Gaussian components larger than 
$30${\arcsec} we do not attempt to find LR matches as the size of these sources 
or components make the LR identification unreliable.

For larger and more complex sources, statistical matching is not reliable so we 
employ a second method for identification and association or separation of 
components. This method involves human visual classification and is built on a 
Zooniverse framework. The project, called LOFAR Galaxy Zoo (LGZ), is described 
in detail in Section  \ref{sec:cat:lgz}. Since it is prohibitive in terms of 
time, as well as unnecessary, to do this for all sources in the \PyBDSF/ 
catalogue, we preselect for LGZ processing samples of sources that are likely to 
be complex.

The sources in the \PyBDSF/ catalogue are selected either for LGZ  processing or 
for acceptance of the LR match based on their catalogued characteristics by 
means of a decision tree described Section\ \ref{sec:cat:flow}. The main 
\PyBDSF/ catalogue parameters we use for the decisions are the source size 
(defined as the major axis), the source flux density, the number of fitted 
Gaussian components, the distance to the NN, and the distance to the fourth 
closest neighbour. In the decision tree we further make use of the LRs 
determined for all sources in the catalogue smaller than $30${\arcsec}, as well 
as the LRs for all the Gaussian components smaller than $30${\arcsec}. The 
thresholds used to determine whether a given source or Gaussian component has an 
acceptable LR match are discussed in Section \ref{sec:cat:ml}.

\vspace{0.5cm}

\section{Likelihood ratio identifications}
\label{sec:cat:ml}
In this Section we describe the statistical LR method and how it is used to  
identify the majority of sources in the LoTSS-DR1 catalogue. The general 
description of the method is given in Section\ \ref{sec:cat:ml_gen} and the 
specific application to the LoTSS-DR1 data set in Section\ \ref{sec:ml:lrapp}. 
As discussed in Section~\ref{sec:data}, deep and wide area data for host galaxy
identifications are available over the LoTSS-DR1 sky area from Pan-STARRS and 
AllWISE. We use a magnitude-only LR method  to cross-match the Pan-STARRS and 
AllWISE catalogues over the LoTSS-DR1 sky coverage and produce a combined 
Pan-STARRS and AllWISE catalogue, which includes sources with detections in only 
PanSTARRS or only AllWISE or both (See Section\ \ref{sec:panstarrswise} for 
details), and thus includes colour information for each source. The LoTSS-DR1 
sources are cross-matched with this combined Pan-STARRS--\textit{WISE} catalogue 
using a colour- and magnitude-dependent LR method (See Section\ \ref{sec:iterq} 
for details).

\subsection{The likelihood ratio method}
\label{sec:cat:ml_gen}
The LR technique (e.g. \citealt{Richter_1975},  \citealt{deRuiter_1977} 
and \citealt{Sutherland_1992}) is a maximum likelihood method used to
statistically investigate whether an object observed at one wavelength
is the correct counterpart of an object observed at a different wavelength. It 
is particularly useful 
when the basis catalogue has a poorer angular resolution or lower source density 
than the catalogue in which the counterpart is being sought, thus giving rise to 
multiple potential matches from which the most likely counterpart needs to be 
identified. This is often the case when seeking optical or IR identifications to 
radio sources, as in this paper. In the description below we specifically use 
`radio' to refer to the basis catalogue and `optical' to refer to the catalogue 
being matched to. However, these terms can be more generally replaced by any 
basis catalogue and matched catalogue -- for example, we also use the LR 
technique to find Pan-STARRS counterparts to AllWISE sources.

The LR of an object is defined as the ratio of the probability of the object 
being
the true counterpart to that of it being a random interloper. This can be
generally written as

\begin{equation}
  LR = \frac{q(x_1,x_2,\dots) f(r)}{n(x_1,x_2,\dots)}
.\end{equation}

\noindent Here, $q(x_1,x_2,\dots)$ represents the \textit{a priori} probability
that the radio source has a counterpart with parameters (which might be
any magnitudes, colours, redshift, type, or any other galaxy property to
be included in the analysis) with values $x_1$, $x_2$,
etc. The parameter $n(x_1,x_2,\dots)$ is the sky surface density of objects with
properties $x_1, x_2$, etc. $f(r)$ is the probability distribution
function for the offset $r$ between the position of the radio source and
its potential counterpart, taking into account the uncertainties in the 
positions of
each. 

Likelihood ratios are commonly calculated using a single galaxy magnitude ($m$)
as the only parameter, in which case 

\begin{equation}
\label{eqn:lr}
LR = \frac{q(m) f(r)}{n(m)}
.\end{equation}

\noindent We use this simple approach 
for cross-matching the PanSTARRS and \textit{WISE} catalogues.  The methods for 
determination of  $f(r)$, $n(m),$ and $q(m)$ are discussed below.

\cite{Nisbet_2018} showed, using an analysis of
LOFAR sources in the ELAIS-N1 field, that including galaxy colour (in
their case, $g-i$ and $i-K$ colours) as well as magnitude greatly
increased the robustness of the  LR analysis for radio source host
galaxies. The inclusion of the $i-K$ colour was particularly useful, as radio 
source 
hosts are well known to be frequently red in optical to near-IR colours:
galaxies of given $i$-band magnitude were found to be around an order of 
magnitude more likely to host a radio source if they had a colour $i-K>4$ than 
those with $i-K<3$. In the LR analysis for the LoTSS sources we therefore 
consider magnitude and colour ($c$), and use

\begin{equation}
\label{lr-eqn}
  LR = \frac{q(m,c) f(r)}{n(m,c)}
.\end{equation}

Specifically, we use the Pan-STARRS $i$-band
data and the \textit{WISE} $W1$ (3.4\,$\mu$m) data, as these 
offer the highest detection fractions for the radio sources and also
provide an optical--to--IR colour baseline similar to the $i-K$
colour used by \cite{Nisbet_2018}.

\subsubsection{Determination of $f(r)$}

The parameter $f(r)$ represents the probability distribution of offset $r$ between 
the
catalogued positions of the radio source and its potential counterpart. The
uncertainty in this offset is calculated by combining the uncertainty on
the radio position, the uncertainty on the optical/IR position, and
the uncertainty on the relative astrometry of the two surveys. It is
important to take into account that radio positional errors are frequently
asymmetric due to an elliptical beam shape, or an extended radio
source. Therefore we need to evaluate radio-optical offsets relative to the 
major and minor axis direction of each source (as opposed to working in the 
RA and Dec directions, which are in general not aligned with the PSF), as well 
as along the direction between the radio source and possible counterpart. The 
parameter 
$f(r)$ is then given by

\begin{equation}
  f(r) = \frac{1}{2\pi\sigma_{\rm maj}\sigma_{\rm min}}
  \rm{exp}\left(\frac{-r^2}{2\sigma^2_{\rm dir}}\right)
\label{fr-eqn}
,\end{equation}

\noindent where $\sigma_{\rm maj}$ and $\sigma_{\rm min}$ are the combined
positional uncertainties along the radio source major and minor axis
directions, and $\sigma_{\rm dir}$ is the combined positional uncertainty
projected along the direction from the radio source to the possible
counterpart under investigation. We now discuss each component of the
positional error budget in turn.

For each LoTSS source, \PyBDSF/ returns the error on the full width at half 
maximum (FWHM) of
the major and minor axes for the fitted Gaussian ($\delta_{\rm
  FWHM,maj}, \delta_{\rm FWHM,min}$) as well as the position
angle. As shown by \cite{Condon_1997}, the uncertainty on the radio position
along the major (minor) axis direction ($\sigma_{\rm maj(min),rad}$) is formally
given by $\sigma_{\rm maj(min),rad} = \delta_{\rm FWHM,maj(min)} /
(8\ln2)^{1/2}$. However, this does not take into account the presence
of correlated noise in the radio images; empirical results from the
NVSS (\citealt{Condon_1998}) and WENSS (\citealt{Rengelink_1997}) surveys
indicate that the formal positional errors on the radio sources are
typically a factor of 1.3--1.5 larger. Here, a factor $\sqrt 2$ is
adopted, and so the positional uncertainties along the major and minor
axes are $\sigma_{\rm maj(min),rad} = \delta_{\rm FWHM,maj(min)} /
(4\ln2)^{1/2}$. Then, using the angle between the major axis direction
and that of the vector joining the LoTSS source to its potential
counterpart, these two uncertainties are projected to derive the radio
positional uncertainty in the direction of the potential counterpart
($\sigma_{\rm dir,rad}$).

The positional uncertainties for the optical/IR galaxy are
catalogued in the RA and Dec directions; these are therefore re-projected
into the radio source major axis, minor axis, and source-to-counterpart 
directions
($\sigma_{\rm maj,opt}$, $\sigma_{\rm min,opt}$ and $\sigma_{\rm
  dir,opt}$), although in practice these uncertainties are often
symmetric. For the astrometric uncertainty between the radio and
counterpart surveys, a value of $\sigma_{\rm ast} = 0.6${\arcsec} is
adopted. This is larger than the typical astrometric uncertainty
determined by \citetalias{Shimwell_2018} but, as discussed in
\cite{Nisbet_2018}, it is important to take a conservative approach as the
astrometric errors are generally not Gaussian. For most sources, the
astrometric uncertainty makes a negligible contribution to the overall
uncertainty, but adoption of too small a value can lead to a failure to
select genuine counterparts for some bright compact radio sources for
which signal-to-noise dependent positional uncertainties can be
unrealistically small. The value of $\sigma_{\rm ast} = 0.6${\arcsec} was 
chosen empirically by visually examining borderline cases of bright compact 
radio sources.

These three contributions are combined in quadrature to derive the overall
positional uncertainty required in Equation~\ref{fr-eqn}, i.e.       

\begin{equation}
  \sigma^2_{\rm maj} = \sigma^2_{\rm maj,rad} + \sigma^2_{\rm maj,opt} +
                      \sigma^2_{\rm ast} 
\end{equation}

\noindent and similarly for $\sigma_{\rm min}$ and $\sigma_{\rm
  dir}$. Thus, $f(r)$ can be calculated for each potential
counterpart.

\subsubsection{Determination of $n(m)$ and $n(m,c)$} 

The parameter $n(m)$ represents the number of objects per unit area of sky at a 
given magnitude, and is easily calculated using a well-defined, representative 
large region of sky,
which is not significantly affected by bright stars or other limitations
that cause incompleteness in the survey. A Gaussian kernel density estimator 
(KDE) of width 0.5~mag was used to determine $n(m)$; particularly for the 
smaller
number statistics of $q(m)$ at bluer colours (see Section~\ref{sec:iterq}), a 
KDE provides smoother and more
robust results than binning. 

In colour space, to determine $n(m,c)$, the sample is divided
into colour bins and $n(m)$ is determined separately for galaxies within
each colour bin. Adoption of a two-dimensional KDE in both
colour and magnitude was considered, but would have required highly
adaptive scaling lengths to account for both the broad colour tails and
the rapid changes in $q(m)/n(m)$ at intermediate colours.

\subsubsection{Determination of $q(m)$} 
\label{sec:cat:ml_q}

The parameter $q(m)$ represents the {\it a priori} probability that the radio source 
has a
counterpart of magnitude $m$. Ideally this would be
predetermined using an independent data set. However, in general this is
not possible and the data set itself must be used; great care must be
taken to avoid biases due to galaxy clustering.

Methods to estimate $q(m)$ have been developed by \cite{Ciliegi_2003},
\cite{Fleuren_2012}, and \cite{McAlpine_2012}, amongst others. By defining
a fixed search radius $r_{\rm max}$ (typically chosen to be comparable to the 
angular 
resolution of the basis survey), the magnitude distribution of all
optical/IR sources within $r_{\rm max}$ of all the radio sources can be 
determined
(usually referred to as ${\rm total}(m)$). This can be statistically corrected 
for
background galaxy counts to determine the magnitude distribution of just the 
galaxy counts associated with the radio sources (${\rm real}(m)$) using

\begin{equation}
\label{eqn:realm}
  \mathrm{real}(m) = \mathrm{total}(m) - n(m) N_{\rm radio} \pi r^2_{\rm max}
,\end{equation}

\noindent where $N_{\rm radio}$ is the number of radio sources in the
catalogue (and hence the second term accounts for the total sky area out 
to $r_{\rm max}$ around all $N_{\rm radio}$ sources). Determined in this way, 
$\mathrm{real}(m)$ contains the true radio
source host galaxies, but may also include additional galaxies within
$r_{\rm max}$ around the radio sources that are not themselves the host,
but are associated with it (e.g. because radio-loud AGN often
lie in overdense group or cluster environments, e.g. \citealt{Prestage_1988},
\citealt{Hill_1991} and \citealt{Best_2004}). This issue will be
returned to shortly.

The parameter $q(m)$ is then derived from ${\rm real}(m)$ as

\begin{equation}
  q(m) = \frac{\mathrm{real}(m)}{\sum_{m_i} \mathrm{real}(m_i)} Q_0
,\end{equation}

\noindent where $Q_0$ represents the fraction of sources that have a
counterpart down to the magnitude limit of the survey (i.e. $Q_0 = N_{\rm 
matched} / N_{\rm radio}$). \cite{Fleuren_2012}
outlined a method to derive $Q_0$ in a manner unbiased by galaxy
clustering by comparing the number of the fields around the radio sources
which are blank (i.e. without any possible counterparts) out to a chosen search 
radius\footnote{In theory the resultant $Q_0$ should be insensitive to the 
radius chosen. In practice, $Q_0$ is usually evaluated for a range of radii 
around the angular resolution of the basis catalogue, and an average value 
taken.} $r_s$, (referred to as $N_{\rm blank}(r_s)$) to the number of blanks 
around an equivalent number of randomly chosen positions ($N_{\rm 
blank,ran}(r_s))$,  
\begin{equation}
\label{eqn:Q0}
  F(r_s) Q_0 = 1 - \frac{N_{\rm blank}(r_s)}{N_{\rm blank,ran}(r_s)}
,\end{equation}

\noindent where $F(r_s)$ is the fraction of the true identifications that
are expected to be found within radius $r_s$. Formally $F(r_s)$ should be 
derived by
integrating $f(r)$ for each source, across all position angles, out to $r_s$, 
but in
practice it is accurate enough to take an average value of $\sigma$, in
which case $F(r_s) = 1 - {\rm exp}(-r_s^2 / 2\sigma^2)$.

Derived in this way, $Q_0$ is unbiased by the effects of galaxy
clustering; this is because the calculation relies on counting
blank fields, so is unaffected by whether a detected radio source host galaxy 
also 
has associated companion galaxies within the search radius. However, as noted 
above, 
the magnitude distribution $q(m)$ may still be mildly affected by the companion 
objects.

\subsubsection{Determination of $q(m,c)$} 
\label{sec:cat:ml_qmc}

This same method cannot easily be adopted across
different colour bins. Although $\mathrm{real}(m,c)$ can be easily determined in
each colour bin using Eqn.~\ref{eqn:realm}, the \citeauthor{Fleuren_2012} 
method 
of Eqn.~\ref{eqn:Q0} is not able to correct for clustering biases in the 
determination of  $Q_0(c)$ (the fraction of sources with a counterpart of colour 
$c$, such that $Q_0(c) = N_{\rm matched}(c)/N_{\rm radio}$ and $\sum_c
Q_0(c) = Q_0$). This can be seen by considering the case of a radio source host 
in one colour
bin which has a physically associated galaxy (i.e. a companion galaxy within the 
same group or cluster) within the search radius, but which falls in a different 
colour bin. In this case, as well as (correctly) not being a blank field in the 
colour bin of the true host galaxy, that radio source would also not be a blank 
field when examining the colour bin corresponding to the companion galaxy. Since 
the companion galaxy is not a random interloper, the search around random 
positions ($N_{\rm blank,ran}(r_s)$) would not correct for this. Hence, this 
radio source would contribute towards $Q_0(c)$ in the colour bins of both the 
true host galaxy and the companion, leading to an overestimate of $Q_0$ by as 
much as tens of percent for larger values of $r_s$.

Instead, therefore, we adopt the process developed by \cite{Nisbet_2018}, which 
is to derive $q(m,c)$ through an iterative approach. Our specific adaptation of 
this is outlined in more detail in
Section~\ref{sec:iterq}, but in summary the iterative approach works as follows:

\begin{enumerate}
  \item First, a rough starting estimate is made for the set of host galaxies to 
the radio sources. In principle, this starting estimate could be as simple as a 
NN cross-match out to some fixed radius. In practice, in order to speed up the 
convergence of the iterative procedure, we produce this starting estimate by 
using magnitude-only LR analyses in the Pan-STARRS $i$-band and \textit{WISE} 
$W1$ bands (see Section~\ref{sec:iterq} for the specific details of how we do 
this).
   \item This first-pass list of host galaxies is then split by colour to 
provide a direct estimate of each of the $Q_0(c)$ -- the fraction of radio 
sources which have counterparts within each colour bin. Dividing by magnitude as 
well then gives a first estimate of $q(m,c)$ -- the fraction of radio sources 
with a counterpart of magnitude $m$ and colour $c$.
  \item Using this $q(m,c)$ estimate, LRs are derived for all galaxies around 
the radio sources (out to some radius -- in our case $15\arcsec$) using both 
magnitude and colour parameters.
  \item Using these LR values, a revised estimate for the list of host galaxies 
is produced by selecting the highest LR match to each radio source, provided 
that it exceeds the LR threshold (see Section~\ref{lr_comprelsec}).
  \item This revised set of matches is used to provide improved estimates of 
$Q_0(c)$ and $q(m,c)$, and steps 3 to 5 are iterated to convergence. 
\end{enumerate}

\subsubsection{Likelihood ratio thresholds}
\label{lr_comprelsec}

Once all three probability distributions ($f(r)$, $n(m)$ and $q(m)$, or $n(m,c)$ 
and $q(m,c)$) are determined, Equation~\ref{eqn:lr} or~\ref{lr-eqn} (as 
appropriate) can be
used to determine the LR of each candidate host
galaxy. The remaining issue is then to decide which identifications to
adopt. An advantage of the LR technique is that, in
ambiguous cases, multiple possible host galaxy identifications can be
retained, with a probability of association assigned to each. However, for this 
first
LoTSS data release, we retain only the most likely match (i.e. the object with 
the highest LR), if its LR is above our defined threshold level.

For a given LR threshold $L_{\rm thr}$, the completeness
($C(L_{\rm thr})$: the fraction of real identifications which are
accepted) and the reliability ($R(L_{\rm thr})$: the fraction of accepted
identifications which are correct)\footnote{We note that defining the
  reliability in this sense -- referring to the whole catalogue -- is
  distinct from the reliability as used in the LR formalism by
for example\ \cite{Sutherland_1992}.} of the resultant sample can be
determined as (e.g. \citealt{deRuiter_1977} and \citealt{Best_2003})

\begin{equation}
  C(L_{\rm thr}) = 1-\frac{1}{Q_0 N_{\rm radio}} \sum_{LR_i < L_{\rm thr}}
  \frac{Q_0 \, LR_i}{Q_0 \, LR_i + (1 - Q_0)}
, \end{equation}

\begin{equation}
  R(L_{\rm thr}) = 1-\frac{1}{Q_0 N_{\rm radio}} \sum_{LR_i \ge L_{\rm thr}}
  \frac{1 - Q_0}{Q_0\, LR_i + (1 - Q_0)}
,\end{equation}

\noindent where the summation for the completeness calculation is 
over the highest LR counterparts to all sources for which the 
best match has a LR below the threshold, and the summation for the
reliability is for the best matches above the threshold. The choice of
$L_{\rm thr}$ then depends on the relative importance of completeness and
reliability for the sample under investigation, but a typical value might
be where these two functions cross, or where their average is
maximised. We note that the point where completeness and reliability cross is
also the value of $L_{\rm thr}$ which delivers a fraction $Q_0$ of
identifications. This is the threshold adopted for the current analysis.

\subsection{Practical application to the LoTSS data set}
\label{sec:ml:lrapp}

\subsubsection{Combining Pan-STARRS and \textit{WISE} data}
\label{sec:panstarrswise}
Before combining with the radio data, the Pan-STARRS $i$-band and \textit{WISE} 
$W1$-band data sets were first
combined, using a magnitude-only LR analysis. The \textit{WISE}
$W1$ was used as the basis data set and the best Pan-STARRS match (if
any) to each \textit{WISE} source was sought. The matching was done in this
direction, since both the angular resolution and source density of the
Pan-STARRS data are much higher, and so matching in the opposite
direction would lead to multiple Pan-STARRS galaxies selecting the same
\textit{WISE} source. The use of \textit{WISE} data helps the subsequent LR
matching to LoTSS sources given that radio sources are frequently associated 
with galaxies with redder colours and hence brighter near-infrared magnitudes.  
Although we do not explicitly filter out optical galaxies
with no \textit{WISE} emission, our colour-based LR method is effective at 
rejecting these when they are unrelated.

Prior to matching, for the small fraction ($<5\%$) of Pan-STARRS
sources without a measured $i$-band magnitude, the $i$-band magnitude
was estimated from the measurements in the other Pan-STARRS bands
($grzy$) and the mean colours of the all galaxies; this was done by extracting 
the magnitude in each band in which 
the source was detected, adjusting this by the mean colour of all galaxies 
between that band and the $i$-band, and then averaging these values.

Then, using the techniques described above for magnitude-only LRs 
(Section~\ref{sec:cat:ml_gen}) and using the AllWISE catalogue as the basis 
catalogue, an LR
threshold of $L_{\rm thr} = 6.4$ and a value of $Q_0 = 0.62$
were derived (i.e. 62\% of \textit{WISE} $W1$ sources have a counterpart in the
Pan-STARRS $i$-band data). LRs were then derived for all PanSTARRS 
sources  within $15\arcsec$ of each AllWISE position, and for each AllWISE 
source the highest LR above the threshold (if any) was taken as the PanSTARRS 
counterpart.   The counterparts accepted (those
with $LR > 6.4$) are broadly similar to those that would be selected
by adopting a simple NN radial cross-matching out to $\approx 2${\arcsec},
but with a weak magnitude dependence on the allowable radial offset.

A combined Pan-STARRS--\textit{WISE} catalogue was constructed by including all
accepted cross-matches, but also retaining all \textit{WISE} sources without a
Pan-STARRS match, and supplementing the catalogue with all of the
Pan-STARRS catalogue sources that had not been matched to a \textit{WISE}
source. For all catalogue entries, the magnitudes were converted into AB
magnitudes and corrected for Galactic reddening using the data of 
\cite{Schlegel_1998}. 
The overall catalogue contains around
26.5 million entries, of which just over 30\% had detections in both
bands, nearly 20\% were detected only in \textit{WISE}, and 50\% were detected 
by
Pan-STARRS only. Some issues will undoubtedly remain with the combined
catalogue, for example in cases where two nearby Pan-STARRS sources are
blended in the lower resolution \textit{WISE} data into a single catalogue 
entry;
however, these are sufficiently rare that they are not expected to have a 
significant effect on subsequent LoTSS cross-matching. We note that no attempt 
was made to separate stars from galaxies in 
the combined catalogue: LoTSS sources may match to stellar objects (either 
genuine -- such as Pulsars -- or misclassified objects such as quasars) and the 
adopted colour-dependent procedure already works sufficiently well at 
down-weighting the LRs of stellar candidates that attempting to exclude these 
would introduce more errors or biases than potential benefit.

\subsubsection{Combining LoTSS and Pan-STARRS--\textit{WISE} data}
\label{sec:iterq}

We use the full colour- and magnitude-dependent LR method described in Section\ 
\ref{sec:cat:ml_gen} to cross-match the LoTSS-DR1 sources with the combined 
Pan-STARRS--\textit{WISE} catalogue. Specifically, in the LR analysis we 
consider the $i$-band magnitude ($m$) and  the $i-W1$ colour ($c$). For the 80\% 
of sources with detections in Pan-STARRS, we use the Pan-STARRS positions, while 
for the remainder we use the \textit{WISE} positions.

From within the overall LoTSS-DR1 sample, the subset of radio sources for which
LR analysis is appropriate was selected. These are ideally the sources for which 
the \PyBDSF/ radio source position provides a well-defined location for where 
the radio source host galaxy is expected to be, and not those \PyBDSF/ sources 
that are parts of a larger source or are very significantly extended and thus 
have poorly defined positions. Initially, for  this sample we included all LoTSS 
sources smaller than $30\arcsec$. This initial sample was used to calibrate the 
$q(m,c)$ values and calculate the LRs as described in this section, noting that 
these values and LRs are slightly biased by the inclusion of some sources for 
which LR analysis is not appropriate. The full decision tree, using the LRs as 
described in Section~\ref{sec:cat:flow}, was then used to reselect the sample of 
LoTSS sources for which LR analysis is appropriate. We also excluded any 
\PyBDSF/ source already associated in LGZ. This cleaner sample was later used to 
recalibrate the $q(m,c)$ values, recalculate the LRs, and hence derive the 
cross-matched counterparts.

As a starting point for the iterative procedure to derive  $q(m,c)$ described 
above (Section\ \ref{sec:cat:ml_qmc}), an initial pass of determining optical/IR 
counterparts is required. This was achieved by cross-matching the radio sources 
selected for LR analysis against the $i$-band and $W1$-band catalogues 
separately, in each case using a LR analysis considering magnitude only. 
Specifically, for this magnitude-only matching, first the \cite{Fleuren_2012} 
technique was used to derive values of $Q_{\rm 0,i} = 0.512$ and $Q_{\rm 0,W1} = 
0.700$ (i.e.\ 51\% and 70\% identification rates for LoTSS sources in
the $i$ and $W1$ bands, respectively) and the corresponding $q(m)$ 
distributions. Then, the LRs were then derived for all sources in each of the 
$i$-band and $W1$-band catalogues located within $15\arcsec$ of each radio 
position. Sources were accepted as matches if their LRs were above the 
thresholds of $L_{\rm thr} = 4.85$ in the $i$-band or $L_{\rm thr} = 0.70$ in 
the $W1$-band (corresponding to a fraction of $Q_0$ accepted matches in each 
band; see Section~\ref{lr_comprelsec}). If more than one potential counterpart 
was above those thresholds then the counterpart with the highest LR in either of 
the two bands was accepted and the other discarded. Creating the starting sample 
in this manner, rather than a simple cross-match or a LR analysis in one band 
alone, produced a more accurate starting estimate for $q(m,c)$ and led to faster 
convergence of the iterative procedure.

The sources in the combined Pan-STARRS--\textit{WISE} catalogue were then 
divided into 16 colour bins. Two colour bins
corresponded to those objects detected only in the $i$-band and only in the  
$W1$-band. A further 14 colour categories were
defined in $i-W1$ colour for those objects detected in both
bands. These colour categories are detailed in Table~\ref{lr_coltab}.
For each colour category, $n(m,c)$ was determined from the overall 
Pan-STARRS-\textit{WISE} sample. The first-pass LR matches derived above were 
divided by colour and magnitude to
provide the starting estimates of $q(m,c)$ and $Q_0(c)$. 

These values were
then used as the input to a LR analysis using both magnitude
and colour, as per Equation~\ref{lr-eqn}. Specifically, for this analysis, the 
$i$-band magnitude was used to determine the LRs within each colour bin, except 
for the `\textit{WISE}-only' sources for which the $W1$
magnitude was used. As before, the (now colour-based) LRs were calculated for 
all sources in the combined Pan-STARRS--\textit{WISE} catalogue within 
$15\arcsec$ of each radio source position. 

\begin{figure}
\centering
  \includegraphics[width=0.5\textwidth]{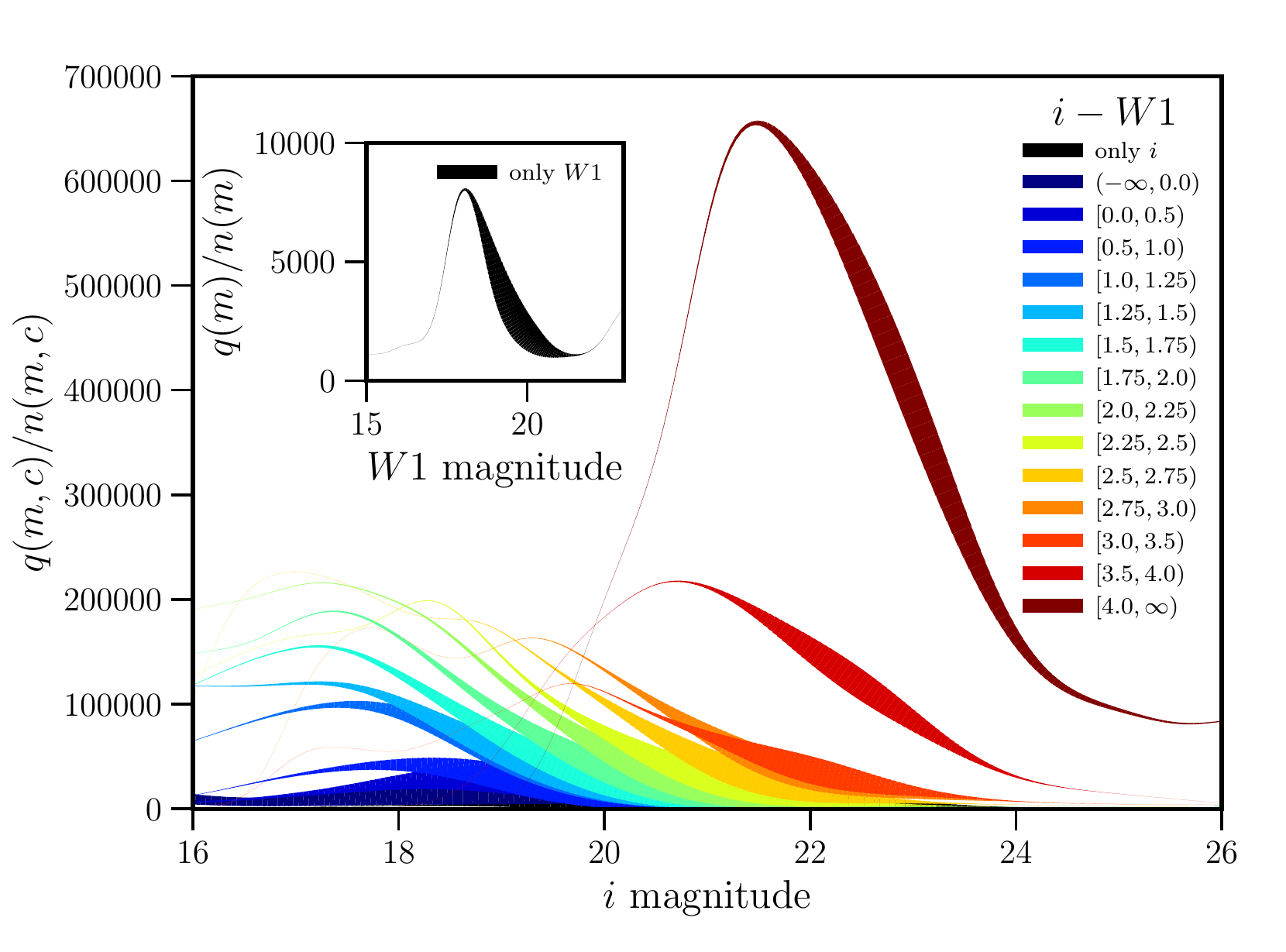}
  \caption{\label{fig_qovern} Plots of $q(m,c)/n(m,c)$ for each colour bin of
    the LR analysis. Lines are colour-coded by galaxy colour bin (running 
naturally from blue to red); the width of the line is proportional to the number 
of LoTSS matches at that magnitude, i.e. thicker regions represent the most 
important regions for $q(m,c)/n(m,c)$ to be determined. The figure clearly 
demonstrates that the KDE approach for
    calculating $q(m,c)$ and $n(m,c)$ is able to produce broadly smooth
    versions of these functions with sufficient magnitude resolution. At fainter 
magnitudes, the ratio $q(m,c)/n(m,c)$ can be seen to rise monotonically and 
strongly towards redder colour bins, i.e. redder galaxies have a higher 
probability to host a radio source, as expected, except at the very brightest 
magnitudes where nearby star-forming (blue) galaxies contribute significantly.}
\end{figure}

From the resultant LRs of the most likely match to each radio source, the LR 
threshold
corresponding to accepting a fraction $Q_0 = \sum_c Q_0(c)$ of
identifications was adopted. The sources with $LR > L_{\rm thr}$ then
provided a modified set of matches, which was used to re-derive $q(m,c)$.
The LRs of all of the Pan-STARRS--\textit{WISE} sources were then re-evaluated 
using the new 
$q(m,c),$ which may lead to a change in the best-matching source or to a source 
moving above or below the LR threshold, and the process was iterated until an 
additional cycle provided no change
in the adopted matches. This required five iterations, although the number of
changes beyond the second iteration was largely negligible. We 
note that in order 
to avoid any risk of systematic bias against the rarest colour categories, a 
minimum value of 0.001 was set for each $Q_0(c)$; the iterative procedure could 
potentially cause $Q_0(c)$ to trend progressively towards zero. 
The final determined values of $Q_0(c)$ are provided in Table~\ref{lr_coltab};
summing these indicates that the total LR identification
rate for LoTSS sources is 73.7\%. The derived $q(m)/n(m)$ functions in
each colour bin are displayed in Fig.~\ref{fig_qovern}.

\begin{table}
  \caption{\label{lr_coltab} Colour bins adopted for LR analysis. The columns 
provide the details of the colour bin (magnitudes are in AB magnitudes), the 
fraction of the combined Pan-STARRS-\textit{WISE} catalogue within that colour 
bin ($f_{\rm PS-\textit{WISE}}$), the iterated value of $Q_0(c)$, the final 
total number of LoTSS source matches to host galaxies of that colour ($N_{\rm 
LoTSS}$) and the fraction of optical/IR sources in the 
combined Pan-STARRS-\textit{WISE} catalogue of that colour that are a match to a 
LoTSS source down to the flux density limit of LoTSS ($f_{\rm radio}$). We note that $N_{\rm 
LoTSS}$ include LR matches to sources included in LGZ associations as explained in Section\ \ref{sec:assoc}, which amount to an average of 2\% of the matches in each bin. }
\small
  \begin{tabular}{ccccc}
    \hline
    \hline
    Colour bin & $f_{\rm PS-\textit{WISE}}$ & $Q_0(c)$ & $N_{\rm LoTSS}$ & 
$f_{\rm radio}$ \\
    \hline
    $i-W1 \le 0$          & 0.034 & 0.0010 &   299 & 0.001 \\
    $0 < i-W1 \le 0.5$    & 0.024 & 0.0056 &  1675 & 0.006 \\
    $0.5 < i-W1 \le 1.0$  & 0.036 & 0.0251 &  6878 & 0.019 \\
    $1.0 < i-W1 \le 1.25$ & 0.026 & 0.0359 &  9459 & 0.037 \\
    $1.25 < i-W1 \le 1.5$ & 0.030 & 0.0514 & 14655 & 0.045 \\
    $1.5 < i-W1 \le 1.75$ & 0.032 & 0.0574 & 16977 & 0.048 \\
    $1.75 < i-W1 \le 2.0$ & 0.031 & 0.0553 & 16885 & 0.047 \\
    $2.0 < i-W1 \le 2.25$ & 0.028 & 0.0500 & 15867 & 0.047 \\
    $2.25 < i-W1 \le 2.5$ & 0.023 & 0.0479 & 14690 & 0.055 \\
    $2.5 < i-W1 \le 2.75$ & 0.017 & 0.0422 & 12813 & 0.063 \\
    $2.75 < i-W1 \le 3.0$ & 0.012 & 0.0362 & 10959 & 0.076 \\
    $3.0 < i-W1 \le 3.5$  & 0.013 & 0.0482 & 14336 & 0.097 \\
    $3.5 < i-W1 \le 4.0$  & 0.004 & 0.0183 &  5429 & 0.120 \\
    $i-W1 > 4.0$          & 0.002 & 0.0059 &  1846 & 0.100 \\
    $i$-band only         & 0.500 & 0.0409 & 11841 & 0.002 \\
    $W1$-band only        & 0.188 & 0.2146 & 65658 & 0.030 \\
    \hline
    Total                 & 1.000 & 0.737 &  220267 &  \\     
  \end{tabular}
\end{table}

\begin{figure}
\centering
  \includegraphics[width=0.5\textwidth]{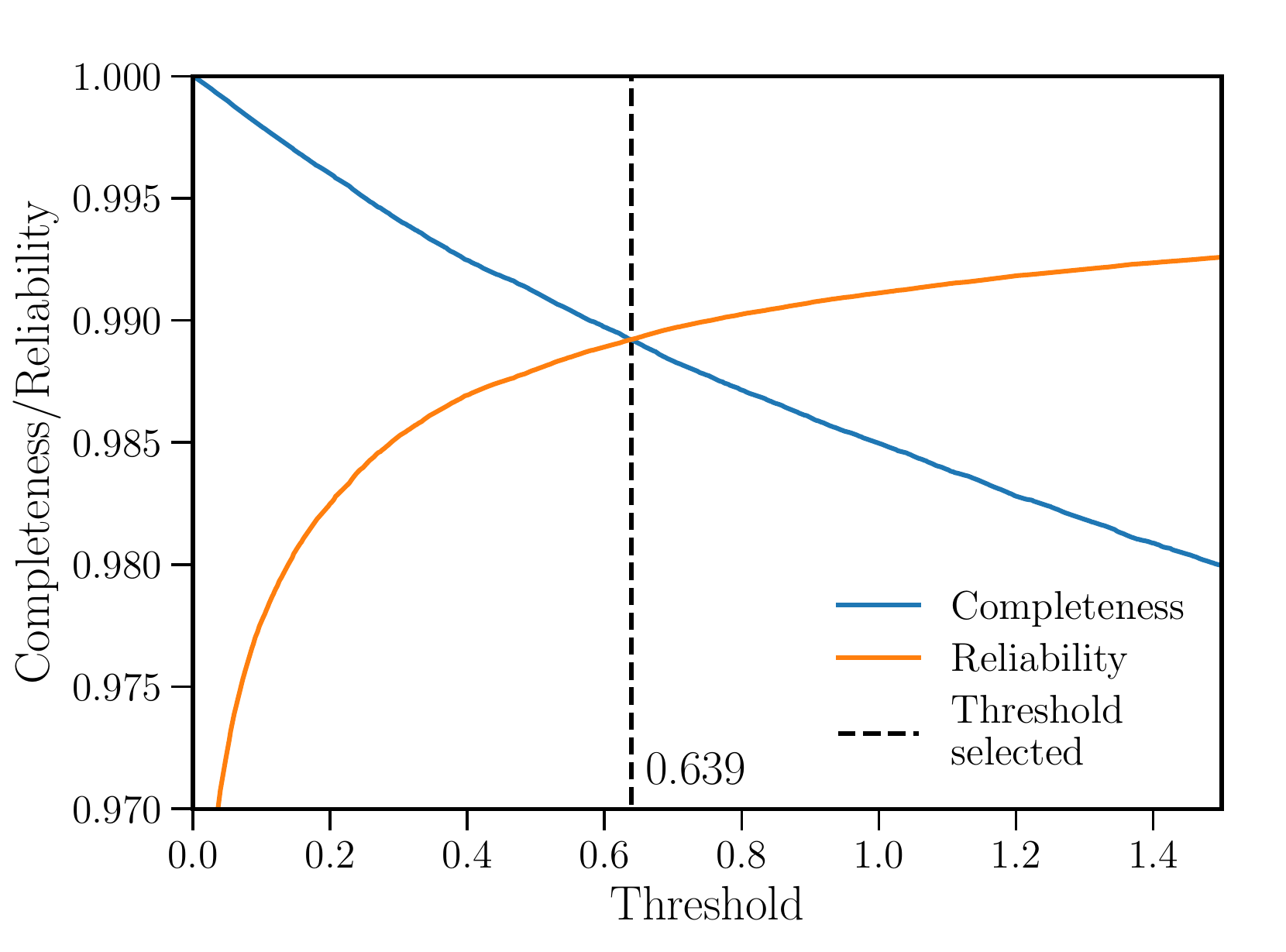}
  \caption{\label{fig_lrthresh} Completeness and reliability of the
    host galaxy identifications as a function of the
    LR threshold. A threshold value of $L_{\rm thr} = 0.639$ 
    was adopted, corresponding to the point where the completeness and 
reliability cross.}
\end{figure}

Final LRs were calculated using the iterated $q(m,c)$.
A plot of the completeness and reliability of the final sample, as a
function of LR threshold, is shown in
Fig.~\ref{fig_lrthresh}. A threshold value of $L_{\rm thr} = 0.639$ that 
corresponds to the point where the completeness and reliability cross was 
adopted (see Section~\ref{lr_comprelsec}). Both the completeness and the 
reliability are $\approx 99$\%.

\begin{figure}
\centering
\includegraphics[width=0.5\textwidth]{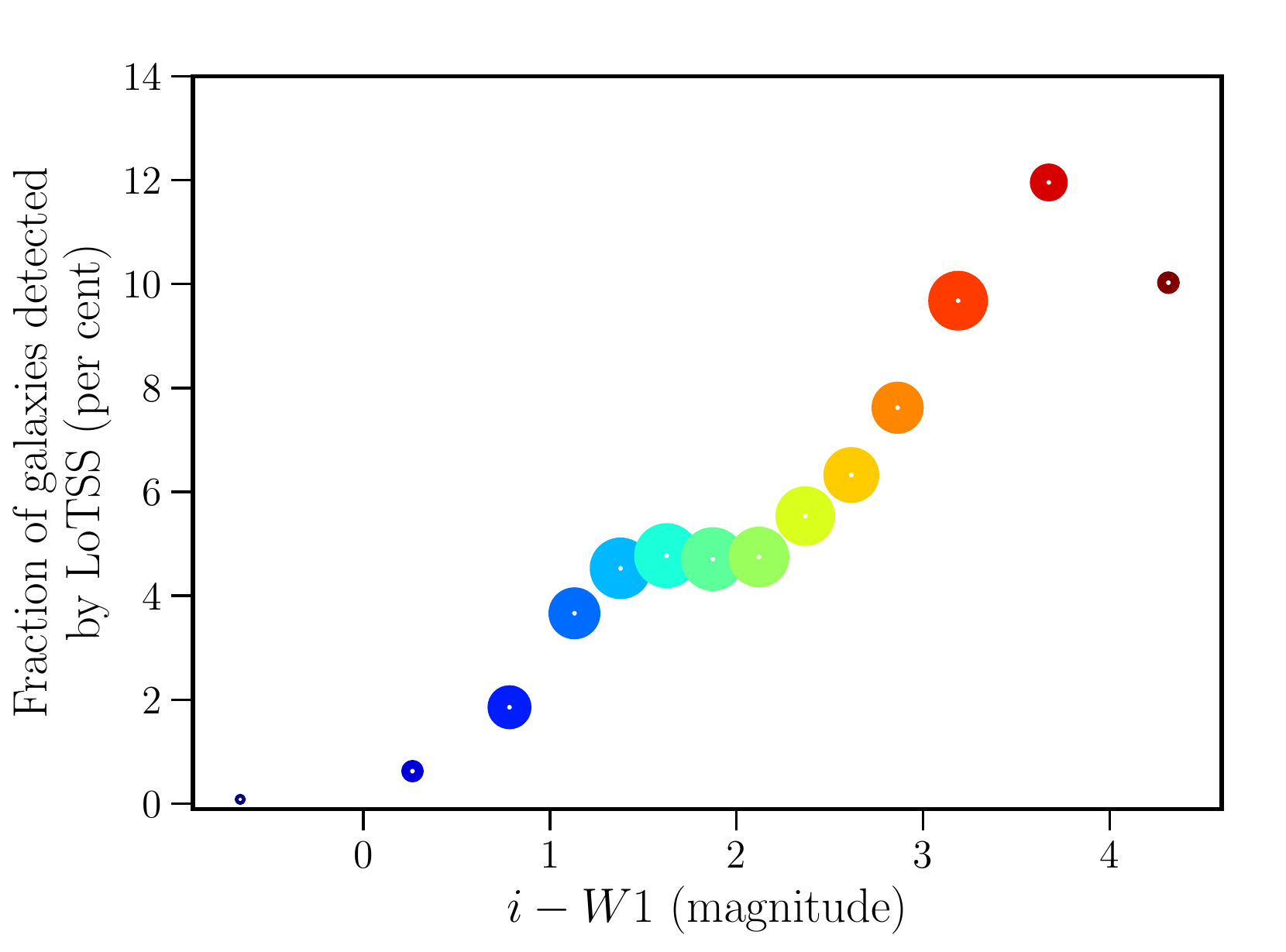}
  \caption{\label{fig_fracrad} Fraction of all galaxies within a particular 
colour bin that have a LoTSS counterpart down to the flux density limit of 
LoTSS. The colour of the symbols corresponds with the colour used in 
Fig.~\ref{fig_qovern}. The position along the x-axis is given by the average 
colour of all the sources in each bin. Poisson error is negligible and the error 
is dominated by misclassification and incompleteness. The size of the marker is 
proportional to the number of LoTSS sources matched. This plot demonstrates the 
additional power of using colour in the LR analysis owing to the much higher 
probability for red ($i-W1>3$) galaxies to host a radio source than for blue 
($i-W1<2$) galaxies to do so.}
\end{figure}

Table~\ref{lr_coltab} shows the number of accepted matches to LoTSS
sources as a function of colour bin. It also shows the fraction of all
galaxies within that colour bin that have a LoTSS counterpart, down to
the flux density limit of LoTSS. This is also shown graphically in
Fig.~\ref{fig_fracrad}, and offers further motivation for the use of
the colour-based LR analysis, since the probability
of the reddest galaxies to host a radio source is an order of
magnitude higher than those of the bluest galaxies.

Now that this has been determined for each colour bin, it can be
applied to any further sample with properties similar to LoTSS. In
particular, it can be used for LR analysis of new survey
areas covered by LoTSS without need for new iterative calculation. We have also 
used this calibrated $q(m,c)$ to derive LRs for counterparts around the 
positions of the individual Gaussian components of
multi-component \PyBDSF/ sources, i.e. for each Gaussian component in the 
\PyBDSF/ Gaussian catalogue, using the \PyBDSF/ Gaussian catalogue as the basis 
catalogue (see also Section~\ref{sec:cat:flow:m}).

\section{Visual identification and association with LGZ}
\label{sec:cat:lgz}

Some sources are too large or complex to be reliably identified through the 
statistical LR technique described in the previous section. Moreover, the LR 
method cannot identify and correct cases where the source finder has not 
correctly grouped components of a single physical source together  or where it 
has incorrectly grouped (blended) multiple physical sources together. Such 
association or deblending needs to be done separately; we do this and the 
optical/IR identification of large and complex sources through visual 
inspection. Based on the properties of the radio sources, we selected a 
subsample of sources to be handled this way; the details of the decision process 
are given in Section\ \ref{sec:cat:flow}. In total, we selected around 13,000 
\PyBDSF/ sources that
plausibly require visual inspection for optical/IR identification or
source association.   

In pilot projects we carried out this sort of
process using manual tools that involved visual inspection of data
stored on a local server by one or a few individuals
\citep{Williams_2016,Hardcastle_2016}; but this is impractical for the HETDEX 
field and still
more so for the larger sky areas that will be provided by the full LoTSS
survey. Instead we used the Zooniverse\footnote{\url{www.zooniverse.org}} 
framework
 and in particular the {\sc panoptes} project
 builder\footnote{\url{https://github.com/zooniverse/Panoptes}} to
 create an association and identification tool which we call
LGZ and which is described in this section. At
this stage of the LoTSS survey, access to LGZ through the web interface was 
limited to members
of the LOFAR Surveys Key Science Project (KSP) and some of their close 
associates. Therefore although we use
the standard Zooniverse terminology and describe the participants in
the project as `volunteers' in what follows, it should be borne in
mind that this is not citizen science and our volunteers all have some
background in professional astronomy. The LGZ project should not be
confused with the very similar Radio Galaxy Zoo project
\citep{Banfield_2015}, from which it draws some inspiration and which
is a true citizen science project. Radio Galaxy Zoo itself is modelled
on the original `Galaxy Zoo' \citep{Lintott_2008} project, which very 
successfully used citizen scientists to classify the morphologies of millions of 
galaxies in SDSS.

\subsection{The LGZ interface}

As in our pilot projects, we made the design decision to carry out in parallel 
the two processes of `association' (where the volunteer decides whether several 
sources in the \PyBDSF/ catalogue should be treated as a single source) and 
`identification' (where the volunteer selects zero, one or more optical host 
galaxies 
for the possibly associated radio source). In many cases the position of a 
plausible optical host is very helpful in deciding on the correct source 
association, or vice versa. We therefore needed to present the volunteer with 
images to classify that showed the radio data and at least one optical image. 
After some experimentation, we chose to use both the Pan-STARRS $r$-band image 
and {\it WISE} band 1, together with radio contours from both the LoTSS images 
and the FIRST survey. The FIRST contours are used alongside LoTSS because 
flat-spectrum cores (which will appear strong in both LoTSS and FIRST), if 
present, are useful in pinpointing a host galaxy, though 
of course the majority of our sources have no FIRST counterpart. Pan-STARRS 
$r$-band is used for its good angular resolution; the ID fraction is only 
slightly lower 
than that of the $i$-band and the bluer wavelength provides a longer colour 
baseline.
We use{\it WISE} band 1 because it is the most sensitive optical/IR band 
available 
to us for the typical elliptical hosts of radio-loud AGN (see 
Section~\ref{sec:cat:ml}), 
although its resolution is much lower than that of Pan-STARRS; 
at $6.1${\arcsec} {\it WISE} band 1 is very comparable to the resolution of the 
LoTSS images 
themselves.

\begin{figure*}
   \includegraphics[width=0.33\linewidth]{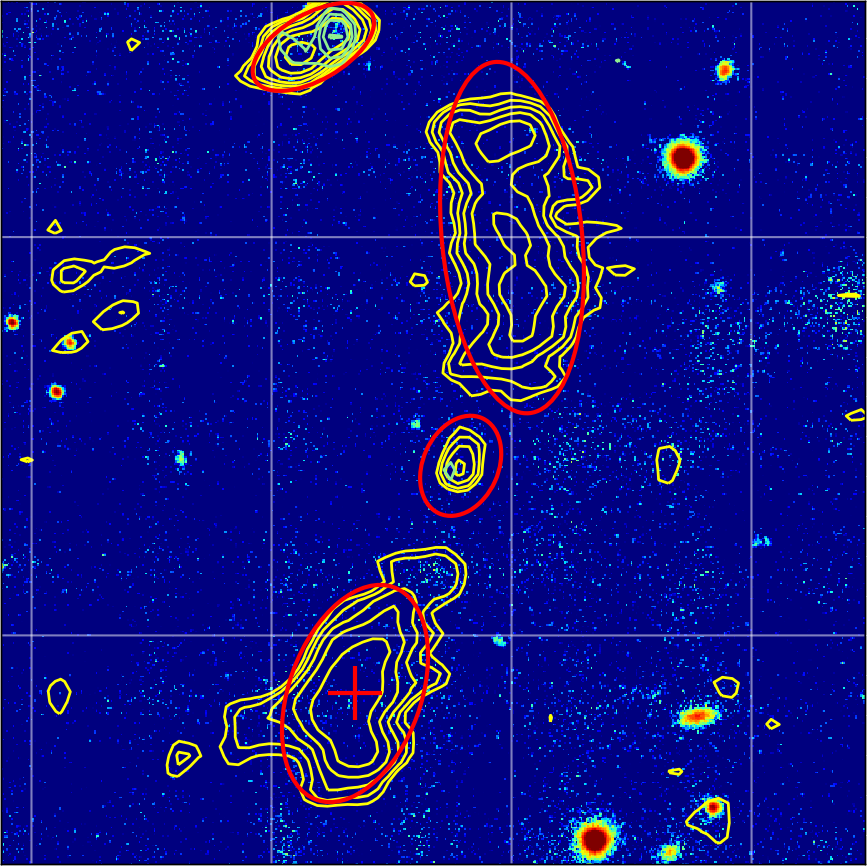}
   \includegraphics[width=0.33\linewidth]{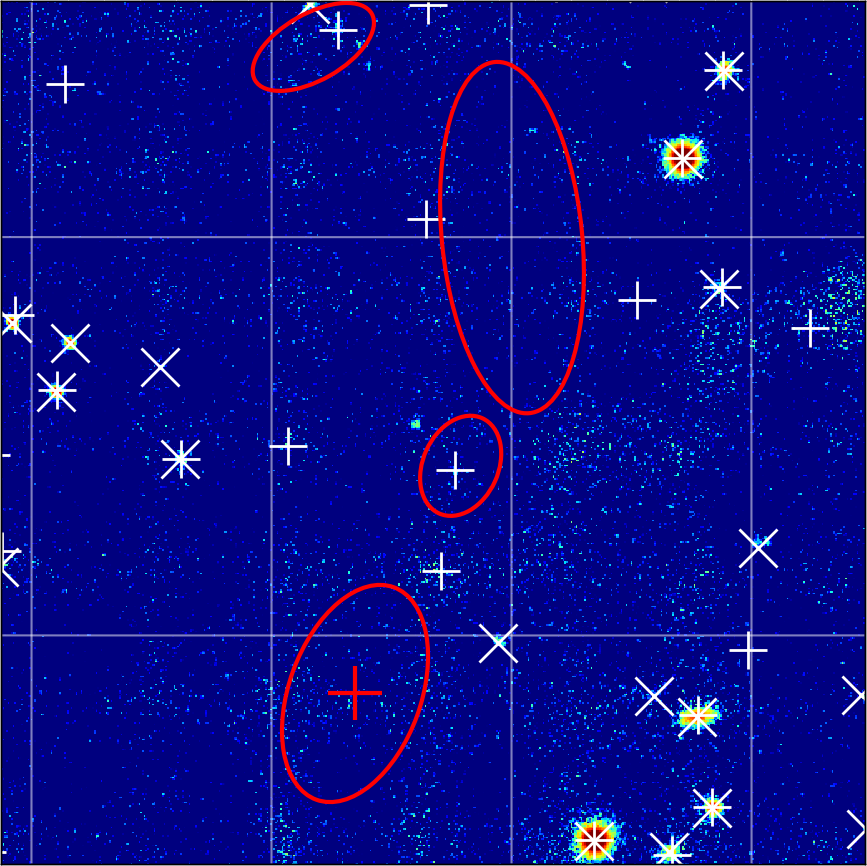}
   \includegraphics[width=0.33\linewidth]{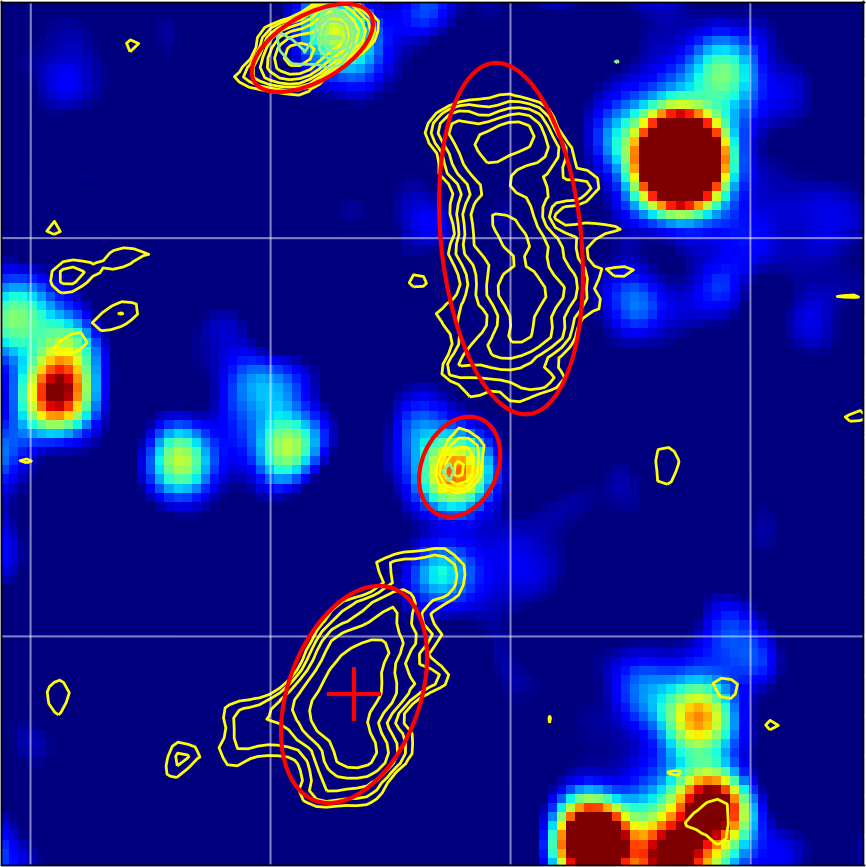}
   
   \includegraphics[width=0.33\linewidth]{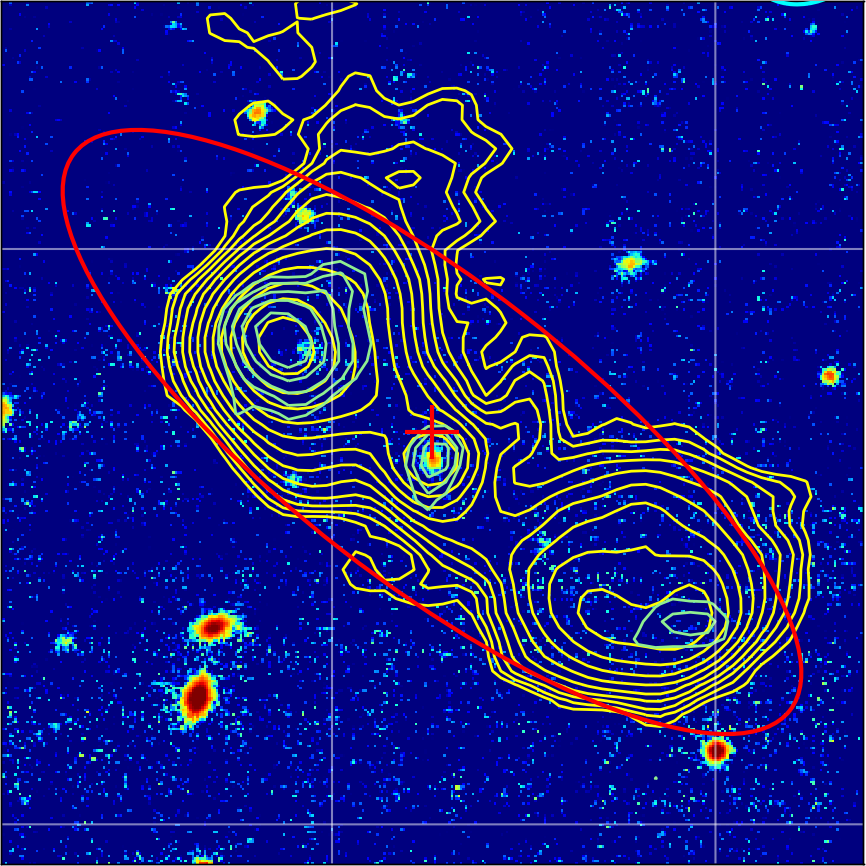}
   \includegraphics[width=0.33\linewidth]{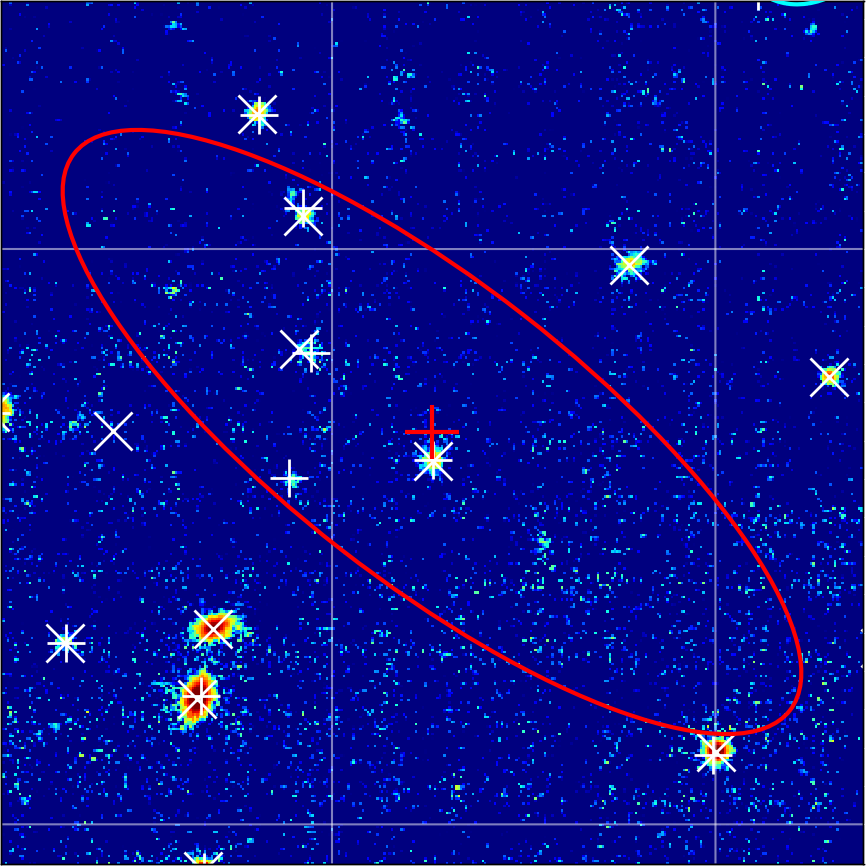}
   \includegraphics[width=0.33\linewidth]{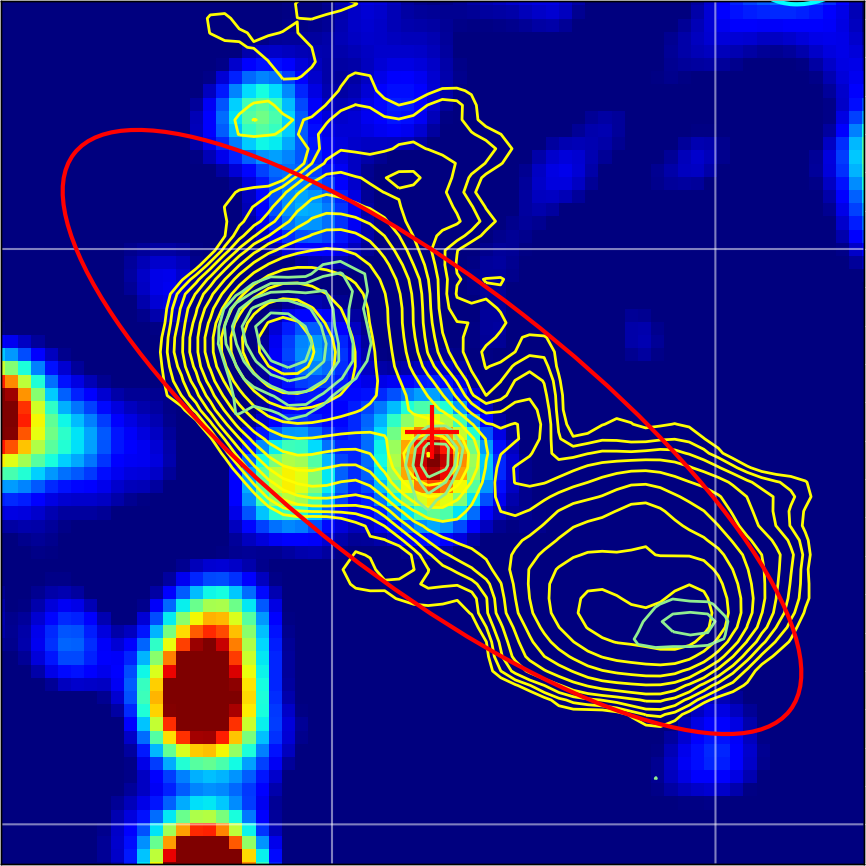}
   \caption{Example set of images from LGZ for two different sources (top and 
bottom). From left to right:
     LoTSS (yellow contours), FIRST (green contours), and Pan-STARRS
     (colour); Pan-STARRS (colour) and Pan-STARRS and \textit{WISE} catalogued
     sources (x's and crosses, respectively); LoTSS, FIRST, and \textit{WISE} 
band 1 (colour). The gridding
     interval in the vertical (N-S) direction is 1 arcmin. In the top example the
     \PyBDSF/ object of interest (indicated with the red cross) is a
     lobe of a radio galaxy. The volunteer should associate it with
     the core and northern lobe, but not with the smaller source on the
   northern edge of the image, which appears unrelated. No Pan-STARRS
   counterpart to the radio source is apparent, but there is a clear
   \textit{WISE} band 1 detection and a marginal FIRST detection (green
   contours) co-located with the central LoTSS component, suggesting
   that this is very probably the host galaxy. In the bottom example there is no 
   other \PyBDSF/ source to associate with the one of interest and there are clear 
   Pan-STARRS and \textit{WISE} detections coincident with the FIRST core.}
   \label{fig:lgz-example}
\end{figure*}

In order to present the images to volunteers in the {\sc panoptes}
framework we have to render them as static images for each \PyBDSF/
source. After trials we settled on three images: one showing LoTSS and
FIRST contours overlaid on a colour scale of the Pan-STARRS $r$-band
image; one with only the $r$-band image, but with catalogued Pan-STARRS
and {\it WISE} sources marked with (distinct) crosses; and one with
the same contours as the first image, but overlaid on a colour scale
of the {\it WISE} band-1 images. All images show ellipses which mark
the location and size of the \PyBDSF/ sources. The {\sc
  panoptes} framework allows the volunteer to flip between these images at any
time, either manually or with automatic cycling, so it is relatively
easy to search for, for example the {\it WISE} counterpart of a Pan-STARRS
source that might be a counterpart to a LoTSS target. Images were made
using the {\sc APLpy} {\sc Python} package \citep{Robitaille_2012};  the
colour and contour levels were determined based on the local image
properties (e.g. local rms noise) and the peak flux density of the LoTSS
source. Specifically, contours were drawn at a lowest level of
  twice the local rms noise level or $1/500$ of the peak flux density of the
  component of interest, whichever was the higher, and increased by a
  factor of 2 from that lowest level. The size of the region to be
displayed was based on both the size of the \PyBDSF/ source of
interest and on the locations of potential association candidates,
using an iterative NN algorithm with some constraints
to prevent the field of view of the image becoming too large or
excluding the original source. Two example image sets are shown in
Fig.~\ref{fig:lgz-example}.

The volunteer can access all three of these images while responding to the
following three sets of instructions:
\begin{enumerate}
  \item Select additional source components that go with the
    LoTSS source marked with the cross. If none, do not select
    anything.
  \item Select all the plausible optical/IR identifications. If
    there is no plausible candidate host galaxy, do not select anything.
  \item Answer the questions: Is this an artefact? Is more than one source 
blended in the
    current ellipse? Is the image too zoomed in to see all the
    components? Is one of the images missing? Is the optical host
    galaxy broken into many optical components?
\end{enumerate}

Answers to these must be provided in order. For tasks (1) and (2) the user 
clicks on the image and the location of their click is stored. For task (3) the 
user checks one or more boxes if the 
answer to the corresponding question is `yes'. The purpose of task (3) is to 
ensure that common problems with the classification are flagged by the user. 
Once all questions are answered, the user can move to the next \PyBDSF/ source.

The Zooniverse interface presents all images to all volunteers until a given 
image 
has been seen a predetermined number of times, after which it is `retired' and 
will no longer 
be presented to volunteers. Originally, we set the retirement limit to ten -- 
that is, 
each image must be classified by ten volunteers before it is retired -- but 
after some 
experimentation we found that we were able to reduce the limit to five in the 
course of the classification process while still recovering good 
classifications. A feature of 
the fact that we present \PyBDSF/ sources to the volunteers is that a complex 
physical source containing a large number of  \PyBDSF/ source components will be 
seen more times than a 
simple one. For example, the top source shown in Fig.~\ref{fig:lgz-example} 
will 
have been seen at least ten times because both the northern and southern lobe 
of 
the radio galaxy meet the selection criterion for visual inspection. We note 
that the \PyBDSF/ source marking the core of the radio galaxy in this example 
would not have been included in the LGZ sample because of its compact nature but 
is included in the output LGZ association. The bottom source in 
Fig.~\ref{fig:lgz-example}  will  only be seen five times.

The LGZ project was carried out in two phases, the first (LGZ v1) was the 
inspection of about 7,000 bright, extended sources in the early part of the 
decision tree (branch A), and the second (LGZ v2) involved around 9,000 later 
decision tree endpoints. 
In LGZ v2 associations from the decision tree and from LGZ v1 were highlighted 
with 
different colours of ellipses and some improvements were made to the code to 
determine field of view, but otherwise there were no significant differences 
between the two parts of the project. One point to note is that LGZ v1 was 
started with an earlier round of processing of the LoTSS images and as a result 
there 
were some differences between the input \PyBDSF/ catalogue for LGZ v1 and the 
final catalogue by the time LGZ was complete. These differences were resolved 
by 
cross-matching of the two catalogues in post-processing and have little effect 
on the final 
results.

\subsection{LGZ output}
\label{sec:cat:lgz:out}

As with all {\sc panoptes} results, LGZ outputs are provided in a JSON file 
which gives details of the location (in pixel terms) of each mouse-click on an 
image and of the answers to the questions asked under task (3) above. These 
raw results were converted to selections of \PyBDSF/ sources and optical 
sources using the underlying catalogues. For the source association, task (1), 
clicks were matched to \PyBDSF/ sources by identifying all sources enclosing
the click position, and then in the case of multiple (overlapping) sources at
the click position, selecting the source whose centre is closest to the click
position. For the optical/IR identifications, task (2), click positions were 
matched to
catalogued galaxies by selecting the nearest galaxy in the combined 
PanSTARRS-WISE catalogue to the click position,
provided the separation distance was less than $1.5\arcsec$. The latter
criterion was applied to exclude a minority of spurious/accidental clicks; this
threshold was optimised using visual inspection. We then looked for consensus 
in both the 
association and identification. 

For each input LGZ source, we considered all sets of \PyBDSF/ sources
associated together by at least one viewer (where a `set' contains one
or more \PyBDSF/ sources), assigning the association set quality
(LGZ\_Assoc\_Qual) to be the fraction of all views of this source
region for which the listed association was chosen as the associated
set. Those associated sets with LGZ\_Assoc\_Qual $> 2/3$ were then
considered as candidate sources for the final catalogue.
Because some sets may be subsets of others, there may be more
  than one set for a given source that meets this threshold; for each
  input source we selected for the final catalogue the largest set
  that included that source and met the quality threshold. In a small
number of cases, resulting from non-optimal image sizes not flagged as
problematic via the LGZ process, peripheral source components (e.g.
small/faint components that were not in the LGZ input sample) ended up
in multiple sets. Such overlaps, which were trivially detected in
the final catalogue by checking for \PyBDSF/ sources that lay in more
than one set, were resolved by visual inspection.

Once the associated sources were finalised, the LGZ optical IDs were
determined in a similar way: all optical/IR identifications made by at
least one viewer were assigned an ID quality (LGZ\_ID\_Qual)
corresponding to the fraction of source views in which this ID was
selected as the correct one. If there was a single ID selected in more
than two-thirds of source views, this was retained for the final catalogue. For 
both the final association of \PyBDSF/ sources and optical IDs, the quality 
flags (corresponding to the fraction of views for which the catalogued outcome 
was selected) were retained in the final catalogue, allowing for more stringent 
cuts to be made in later analysis.

Sources that emerge from LGZ with flags set to indicate that there were a 
significant number of positive answers in task (3) are dealt with in special 
ways. Where a majority (more than 50\%) of volunteers agree in classifying a 
source as an artefact, 
that source is removed entirely from the final catalogue. Several hundred 
dynamic-range artefacts  around bright sources (see 
Section~\ref{sec:cat:flow:art}) were removed in this way. If a 
significant fraction of volunteers (more than 40\%) classed a source as `too 
zoomed in' -- i.e. the 
field of view presented to them was in their opinion not large enough to carry 
out the association or identification correctly -- then that source was 
re-inspected by a single expert using a \textsc{Python}-based interactive tool 
that generates similar images but with 
the ability to pan and zoom, using the volunteers' association as a starting 
point, and new sources (and potentially a revised optical ID,
to be processed in the same way as other LGZ optical IDs) were added to the 
association if necessary. Sources flagged as blends by more than 40\% of viewers 
were examined in the deblending workflow (see Section~\ref{sec:deblend}).
Sources where the host galaxy was flagged as broken up in the optical catalogue 
by more than 50\% of viewers were simply associated with the nearest bright 
optical galaxy from the 2MASX catalogue, as these were confirmed to be 
exclusively associated with optical sources so bright that the PanSTARRS or 
\textit{WISE} cataloguing algorithms had failed. In this case we record the name 
of the 2MASX match, but take the position from the nearest match for that 2MASX 
source in the merged Pan-STARRS/AllWISE catalogue. The flag to indicate that an 
image was missing was hardly used; we inspected visually all four sources where 
more than 50\% of viewers selected this option and verified that they were 
treated appropriately by the default processing.

\subsection{Associated sources}
\label{sec:assoc}

In the following, associated sources refer to those where separate \PyBDSF/ 
sources have been associated and combined into single new physical sources 
either based on the LGZ output or matches with large 
optical galaxies (see Section\ \ref{sec:cat:flow:bright}). The individual 
\PyBDSF/ sources that make up (i.e. are components of) associated sources were 
removed from the final LoTSS-DR1 value-added catalogue and replaced with the 
associated sources, such that the final catalogue should, to the best of our 
ability, contain only true physical radio sources. We note that LGZ associations 
can include \PyBDSF/ sources from other outcomes of the decision tree described 
in Section~\ref{sec:cat:flow}, in which case the LGZ association takes 
precedence.

For all associated sources, we generated the LoTSS source properties and 
populated the relevant table columns (total flux density, size, radio position, 
and radio source name) by combining the  properties of their constituent 
\PyBDSF/ sources (or \PyBDSF/  Gaussian components in the case of blends -- see 
next section). Some of these combinations are obvious but it is worth commenting 
on a few of them. The position of the source was taken to be the flux-weighted 
mean of the positions of each component. For the total flux density, we simply 
summed the total flux densities of each component. Previous work has shown that 
this normally gives a reasonably accurate flux density measurement compared to 
hand-drawn integration regions, as long as \PyBDSF/ has captured all the flux 
density; this is likely to go wrong in for example\ very large diffuse regions 
where \PyBDSF/ fails to distinguish source from background. For each of these 
properties we propogated the errors of the component parameters as appropriate. 
The peak flux density of the associated source was taken to be the maximum value 
of the peak flux densities of the component sources, along with its 
corresponding error. The rms was taken to be the mean value of the rms for the 
component sources. The S\_Code was updated based on the number of Gaussian 
components in the new source; `S' for a single Gaussian component and `M' for 
multiple.

To determine source sizes we used the convex hull around the set of
elliptical Gaussians: the convex hull is the smallest convex shape
that contains all of the ellipses. To construct the convex hull we
represented each component (\PyBDSF/ source or \PyBDSF/  Gaussian as
approprate) as an ellipse, where the deconvolved FWHM major and minor
axes are taken to be, respectively, the semi-major and semi-minor axes
of the ellipse. The convex hull was constructed around all of the
component ellipses using the {\sc shapely} {\sc Python} package. Then
we took the size of the source (`LGZ\_Size') to be the length of the
largest diameter of the convex hull around the set of elliptical
Gaussians; that is, for all points on the convex hull considered
pairwise, we found the maximum vector separation, and took its
magnitude. The source position angle (`LGZ\_PA') was taken to be the
position angle on the sky of that largest diameter vector. For the
source width (`LGZ\_Width') we adopted twice the maximum perpendicular distance 
of points on the convex hull to the largest diameter vector. These definitions 
have the feature that, if applied to a single ellipse, they return the major and 
minor axis of the Gaussian and its position angle. Source sizes determined from 
the maximum distance between components, as in \cite{Hardcastle_2016}, can be 
significant underestimates where the components are extended: the present 
approach is likely to overestimate the true size in general but gives results in 
better agreement with measurements by hand. We do not provide error estimates 
for the shape parameters in the final catalogue.

\subsection{Deblending workflow}
\label{sec:deblend}

Blended sources, either from LGZ or from the `M' source decision tree
(see Section\ \ref{sec:cat:flow:m}), were examined in a specific
deblending workflow involving a \textsc{Python}-based interactive
visual inspection by a single expert.  Each \PyBDSF/ source  was first
split into its Gaussian components as originally fitted by \PyBDSF/. These 
Gaussians were then re-associated as appropriate into new
radio sources and identified with zero or more optical counterparts,
which were handled in exactly the same way as optical
  counterparts found by LGZ. Around 1,500 sources were dealt with in this way.

In the final LoTSS-DR1 value-added catalogue, \PyBDSF/ sources that were 
identified as blends and processed 
in the deblending workflow were  removed and replaced by sources made 
by combining their component Gaussians; they therefore have properties (flux 
densities, sizes, etc.) 
appropriate for associated sources.   The properties of the Gaussian components 
are combined into single sources in the same way that the component \PyBDSF/ 
sources are combined for associated sources as described in  \ref{sec:assoc}, 
except that we use the parameters (total flux density, position, etc.) from the 
\PyBDSF/ Gaussian catalogue. Notably, for the positions and sizes, this is not 
exactly the same  process by which  \PyBDSF/ combines the fitted Gaussians into 
sources, which is based on image moment analysis, but produces comparatively 
similar results.

\section{Decision tree}
\label{sec:cat:flow}

\begin{figure*}
 \centering
 \includegraphics[width=\textwidth,trim={0 2cm 0 2cm},clip]{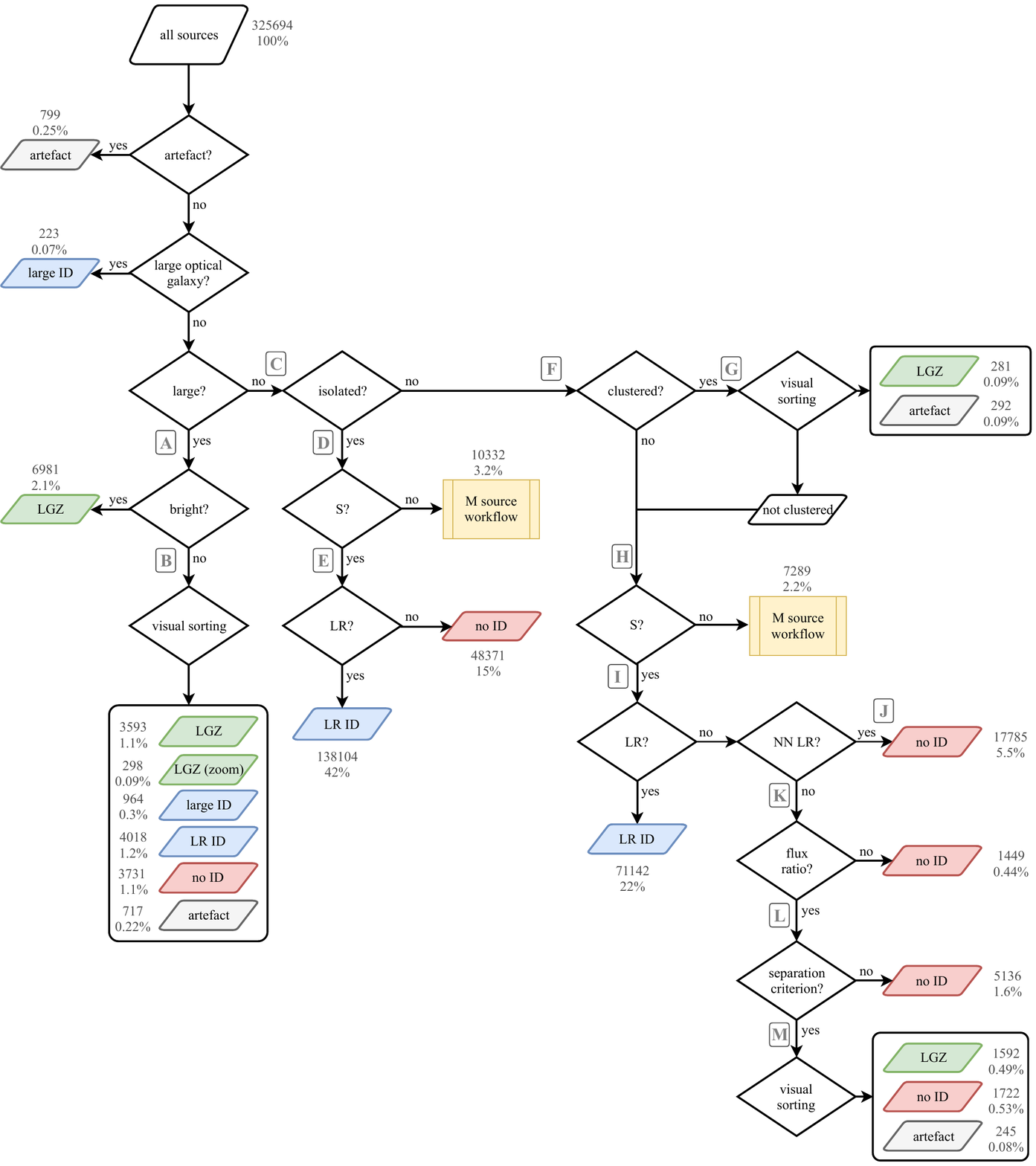}
 \caption{High level summary of the decision tree used to process all entries in 
the  \PyBDSF/ catalogue. Following this workflow a decision is made for each 
source whether to: (i) make the optical/IR identification, or lack thereof, 
through the LR method (blue and red outcomes respectively); (ii) process the 
source in LGZ  (green outcomes); (iii) reject  the source as an artefact (grey 
outcomes); or (iv) process further in a separate workflow (yellow outcomes: see 
Fig.~\ref{fig:cat:flowchartM}). The key parameters are defined in 
Table~\ref{tab:cat:flowkey} and full details of the decisions are given in 
Section\ \ref{sec:cat:flow}, with reference to the branch labels A--M. The 
numbers reflect the number of \PyBDSF/ sources in each final bin and the 
percentage is relative to the total number of sources in the \PyBDSF/ 
catalogue.}
 \label{fig:cat:flowchart}
\end{figure*}

In this section we describe how we select which radio sources to process using 
the statistical LR and visual LGZ methods. We also discuss any sources that need 
to be handled differently. In order to reduce the number of sources that were 
passed to some form of visual inspection, all 325,694 sources in the \PyBDSF/ 
catalogue were evaluated through a decision tree to select subsamples of sources 
that required (i) direct visual association and identification via LGZ; (ii) 
visual sorting into one of several categories, including selection for LGZ;  
(iii) rejection as artefact; or (iv)  identification through LR analysis. We 
describe the main decisions taken, with approximate numbers/fractions of sources 
at each stage. A graphic representation is shown in 
Fig.~\ref{fig:cat:flowchart}, and key parameters are defined in Table\ 
\ref{tab:cat:flowkey} and described in detail in this section. A separate 
process is followed within the decision tree for \PyBDSF/ sources fitted with 
multiple Gaussians. This process is illustrated in 
Fig.~\ref{fig:cat:flowchartM}, and key parameters are defined in Table\ 
\ref{tab:cat:flowkeyM} and described in detail in Section\ \ref{sec:cat:flow:m}. 
These figures and tables are best read as a high-level summary in conjunction 
with the detailed descriptions in the text.

Some stages of the decision tree required `visual sorting' (pre-filtering) prior 
to including sources in the LGZ sample, i.e. to avoid overpopulating the LGZ 
sample with unnecessary sources we filtered them beforehand. For this visual 
sorting, images similar to those used for LGZ (Pan-STARRS $r$-band images with 
radio contours from both the LoTSS images and the FIRST survey) were produced 
and rapidly inspected to categorise the sources relevant to that stage of the 
decision tree. This was done by a small number of experienced people, using a 
simple {\sc Python} interface to view and categorise the images where each source 
was viewed by one person only\footnote{In practice source lists were split 
between several people, each of whom could categorise tens of sources per 
minute.}. The aim of these steps was only to quickly pre-filter the list such 
that the LGZ sample remained manageable and included only the necessary sources; 
i.e. the LGZ sample was not polluted by vast numbers of sources which were 
either clear artefacts or clearly suitable for automated statistical anaylsis. 
The aim was not to also make the LGZ classification as this would slow down the 
process and because visual classifications in LGZ are made by consensus by 
several people.

\begin{table*}[htp]
 \centering
 \caption{Definition of the parameters used in the main decision tree in 
Fig.~\ref{fig:cat:flowchart}. See Section\ \ref{sec:cat:flow} for details.}
 \small
 \label{tab:cat:flowkey}
\begin{tabular}{ll}
\hline
\hline
Parameter & definition \\
\hline
Large optical galaxy & 2MASX size ($r_{\mathrm{ext}}$) $\geq 60${\arcsec} \\
Large & \PyBDSF/ major axis $>15${\arcsec} \\
Bright & total flux density $>10$\,mJy \\
Isolated &  distance to nearest \PyBDSF/ neighbour (NN) $>45${\arcsec} \\
S & single Gaussian component within an island \\
LR &  $LR > 0.639$\\
Clustered & distance to fourth nearest \PyBDSF/ neighbour $<45${\arcsec} \\
NN LR & $LR_{\mathrm{NN}} > 0.639$\\
Flux ratio & $S/S_{\mathrm{NN}} < 10 $\\
Separation criterion & $S + S_{\mathrm{NN}} \leq 50 (d_{\mathrm{NN}} / 
100{\arcsec})^2$\,mJy\\
\hline
\end{tabular}
\end{table*}

\subsection{Artefacts}
\label{sec:cat:flow:art}
Owing to the dynamic range limitations in the imaging \citepalias[see section 
3.4 in][]{Shimwell_2018}, the \PyBDSF/ catalogue contains a not insignificant 
number of spurious sources or artefacts. These are generally found near the 
brightest compact sources in the images. Typically these consist of either 
several small artefacts detected in the vicinity of the bright source, or large 
artefacts in the vicinity of the bright source picked up at the higher order 
wavelet scales of the source detection. Since these are not real sources, they 
need to be flagged as such and removed from the final catalogue.

An initial selection of candidate artefacts was made by considering all compact 
bright sources (brighter than $5$\,mJy and  smaller than $15${\arcsec}) and 
selecting their neighbours within $10${\arcsec} that are $1.5$ times larger. 
This selects large sources in close proximity to compact, bright sources. Since 
such structures can in fact be real, for example faint lobes near a bright radio 
core, these candidate artefacts were visually confirmed. Out of 884 (83\%) of 
such candidate sources 733 were confirmed as artefacts. We note that, as a 
preliminary step, this was not a complete artefact selection; for example it did 
not select clusters of artefacts around bright sources. Further work can be done 
to improve the identification of artefacts at this early stage in the decision 
tree, although future improvements in LOFAR imaging will also reduce the number 
of artefacts. Artefacts were also identified in all further stages of visual 
sorting within the decision tree described here.  Finally, the LGZ output 
included an artefact classification (see Section \ref{sec:cat:lgz:out}). 

Images from pointings on the outer edges of the DR1 coverage have hard edges and 
a small number of sources can be cut off. Sources may still be detected by \PyBDSF/  at 
the edges of an image, but such sources are likely to be incomplete or have 
erroneous flux densities and shapes. We have therefore flagged and removed 
$\sim200$ sources where the fitted \PyBDSF/ shape overlapped the edge of the 
mosaic, or where the source overlapped another edge source.

A total of $2543$ sources ($\sim1$\%) were flagged in  the \PyBDSF/ catalogue 
\citepalias[and an artefact flag column was added to the catalogue presented 
in][]{Shimwell_2018} through the artefact selection and various visual sorting 
and LGZ stages. These sources were dropped from further analysis and are not 
included in the final catalogues presented here.

\subsection{Large optical galaxies}
\label{sec:cat:flow:bright}
The radio emission associated with nearby galaxies that are extended on 
arcminute scales in the optical is clearly resolved in the LoTSS maps and can be 
incorrectly decomposed into as many as several tens of sources in the \PyBDSF/ 
catalogue. To deal with these sources we selected all sources in the 2MASX 
catalogue larger than $60${\arcsec} and for each, searched for all the \PyBDSF/ 
sources that are located (within their errors) within the ellipse defined by the 
2MASX source parameters (using the semi-major axis, `r\_ext', the $K_s$-band 
axis ratio, `k\_ba', and $K_s$-band position angle, `k\_pa'). The \PyBDSF/ 
sources were then automatically associated as a single physical source and 
identified with the 2MASX source. We record the 2MASX source name as the the 
ID\_name of the LoTSS source, but take the co-ordinates and optical/IR 
photometry from the nearest match in the combined Pan-STARRS--AllWISE catalogue, 
with the caveat that the PanSTARRS and AllWISE photometry is likely to be wrong 
for these large sources. This reduced the demands on visual inspection at the 
LGZ stage and avoided the possibility of human volunteers missing out components 
of the radio emission from the galaxy in their classification.

\subsection{Large radio sources}
\label{sec:cat:flow:large}

Since the size of a source is a first indication whether it is resolved and 
possibly complex, we first considered the sources that are large 
($>15${\arcsec}, branch A in Fig.~\ref{fig:cat:flowchart}). This constitutes 
around $6$\% of the sample. All large, bright sources (brighter than $10$\,mJy) 
were selected for visual processing in LGZ\footnote{For the first phase of LGZ 
processing (see Section \ref{sec:cat:lgz}), all large, bright sources in the 
\PyBDSF/ catalogue were selected and so the LGZ v1 sample included some of the 
artefacts and components of large optical galaxies discussed in Sections 
\ref{sec:cat:flow:art} and \ref{sec:cat:flow:bright}}. Containing around 7000 
sources, this constitutes around 2\% of the \PyBDSF/ catalogue.

Instead of also directly processing the remaining $\sim13$k large, faint sources 
(fainter than $10$\,mJy -- branch B) in LGZ, these sources were first visually 
sorted as (i) an artefact; (ii) complex structure to be processed in LGZ; (iii) 
complex structure, where the emission is clearly on very large scales, to be 
processed directly in the LGZ `too zoomed in' post-processing step (see Section\ 
\ref{sec:cat:lgz:out});  (iv) having no possible match; (v) having an acceptable 
LR match, i.e. LR ID; or (vi) associated with an optically bright/large galaxy. 
It should be noted that within this category of large, faint radio sources, 
those larger than $30${\arcsec} are too large to have a LR estimate and so we 
included option (vi) to allow an identification with the nearest large/bright 
optical galaxy based on the Pan-STARRS images. The $\sim1000$ such sources with 
a visually confirmed large optical galaxy match were then matched directly to 
the nearest 2MASX source, or in the 35 cases where there was no 2MASX source, to 
the nearest bright SDSS source. In all cases the nearest 2MASX or SDSS match was 
confirmed to be the correct match. Again the ID positions for these sources are 
taken from the nearest matches in the merged Pan-STARRS/AllWISE catalogue. An 
additional $\sim 4000$ sources were included in the LGZ sample after this visual 
sorting on branch B.

\subsection{Compact radio sources}
\label{sec:cat:flow:compact}

Sources $<15${\arcsec} in size make up around $94$\% of the \PyBDSF/ catalogue 
(branch C). While many of these are individual sources best processed using the 
LR method, a subset are components of complex sources. Visual inspection of the 
entire catalogue was impossible given the available effort, so we applied a 
series of tests to select those small sources most likely to be components of 
complex sources.  We initially considered whether the sources smaller than 
$15${\arcsec} have any nearby neighbours. Sources where the distance to the NN 
is greater than $45${\arcsec}  were considered to be isolated  (branch D; 
$\approx 200$k sources). A separation of $45\arcsec$ corresponds to a linear 
distance of 230--330\ kpc at redshifts of 0.35--0.7, where the bulk of the AGN 
population of this sample is located \citepalias[see][]{Duncan_2018}\footnote{In 
the LOFAR samples of \cite{Hardcastle_2016} and \cite{Williams_2016}, in which 
the association and identification was done entirely visually, $66$\% of the 
sources (i.e. including separate components of AGN) are smaller than 
$45\arcsec$. However this does not mean that we miss larger sources as these are 
picked up in other parts of the decision tree.}. Before directly accepting the 
LR results for these sources, we removed those that were fitted by \PyBDSF/ 
using multiple Gaussian components or those that lay in islands with other 
sources (i.e. with catalogued `S\_Code' values of `M' or `C'); in these cases 
($\approx 10$k sources) a further decision tree was followed, taking into 
account the LR matches to the individual Gaussian components of the source (see 
Section \ref{sec:cat:flow:m}).  For the remaining small, isolated, single 
Gaussian-component sources (i.e. with catalogued `S\_Code' values of `S'), we 
accepted the LR results (branch E): either the source has an acceptable LR match 
(LR ID) or it has no acceptable LR match (no ID).

Small sources that are not isolated (i.e. have at least one other source within 
$45${\arcsec} -- branch F) have a higher chance of being a component of a 
complex source. For these sources we considered whether they are clustered to 
some extent, based on the distance to the fourth neighbouring source: for 
approximately $1100$ sources this distance is less than $45${\arcsec} (branch 
G). Empirically, based on visually examining subsamples of sources, we found 
that taking the fourth NN maximised the number of genuinely clustered sources 
while minimising the number of unrelated sources. As these may be part of a 
larger structure or simply chance groups of unassociated sources that can be 
matched by the LR method, we visually sorted such clustered sources either as 
(i) complex (to be sent to LGZ), (ii) not complex (appropriate for further 
analysis in the decision tree), or (iii) as an artefact. About a quarter of the 
clustered (branch G) sources were selected for LGZ, while about another quarter 
were flagged as artefacts. The remainder were considered not clustered based on 
the visual sorting and  assessed via branch H.

For the remaining small, non-isolated, but not clustered sources (branch H), 
those that have multiple Gaussian components were again treated in a separate 
workflow (see section \ref{sec:cat:flow:m}). We then considered whether the 
source and/or its NN have a LR match above the threshold (branch I).   In the 
case where the source has an LR match, we accepted the LR identification. In the 
case where the source of interest has no LR match, but its NN does, we accepted 
that the source has no match (branch J). However, in the case where neither the 
source nor its NN has an acceptable LR match (branch K), it is increasingly 
likely that the two sources are part of a complex structure where the optical ID 
is not coincident with either radio component. For such pairs, we further 
considered the flux ratio of the source to its NN. Sources with extreme flux 
density ratios are less likely to be associated. We made a somewhat conservative 
cut at a flux density ratio of $10$ \citep[see e.g.][]{Prandoni_2000}, and for 
sources with ratios larger than 10 we accepted that there is no LR match. We 
then applied a flux-dependent separation criterion for the sources with similar 
fluxes (branch L), following \cite{Huynh_2005}, of $S + S_{\mathrm{NN}} \leq K 
(d_{\mathrm{NN}} / 100{\arcsec})^2$, where $S$ and $S_{\mathrm{NN}}$ are the 
total flux density of the source of interest and its NN, respectively, and 
$d_{\mathrm{NN}}$ is their separation in arcsec. The constant $K$  
\citep[$=10$\,mJy in][]{Huynh_2005} was adjusted to take into account the 
different working frequency ($150$\,MHz instead of $1.4$\,GHz). We adopted 
$K=50$\,mJy, under the assumption of steep spectrum radio sources 
($\alpha=-0.7$). For sources that did not meet this criterion we accepted that 
there is no LR match, while for those $\sim3500$ that did (branch M) we did a 
final stage of visual sorting to (i) select as a possible group for LGZ 
association and identification, (ii) accept that there is no match, or (iii) 
classify as artefact. These sources were split roughly equally between the first 
two options and a further $\sim200$ sources were flagged as artefacts.

\subsection{Radio source pairs}
\label{sec:cat:flow:pairs}

The final steps (branches J--M) of the decision tree consider only the NN to a 
given source and not all possible neighbours. To ensure that we did not miss any 
double sources where another unassociated source lies nearer to one of the 
sources than the separation between the pair, we selected all the pairs of 
sources that meet the above flux ratio and flux-dependent separation criteria 
and that also consist of two sources with multiple Gaussian components. To try 
to capture more large radio galaxies, we considered all such pairs with 
separations of up to $60${\arcsec}, not already included in the LGZ 
sample\footnote{Although many giant radio galaxies will be picked up in LGZ, the 
final value-added catalogue may be incomplete for some truly giant radio 
galaxies, in particular those made up of two widely separated compact lobes.}. 
These $\sim3200$ sources were visually sorted and $\sim1500$ ($46$\%) more 
potentially genuine double sources were included in the LGZ sample. Sources not 
included in LGZ keep their classification from the decision tree. This step is 
not shown on the decision tree because it includes sources from several 
different outcomes.

\subsection{Sources with multiple Gaussian components -- `M' sources}
\label{sec:cat:flow:m}

Within the decision tree the largest sources are all visually inspected, either 
directly in LGZ or through visual sorting; however, sources that are small 
($<15${\arcsec}) may still be resolved and may have been fitted by multiple 
Gaussian components by \PyBDSF/. Such sources are identified in the \PyBDSF/ 
catalogue with a value of `M' in the `S\_Code' column, and we refer to these in 
what follows as `M' sources. In this category, we include also the 102 sources 
with `S\_Code' values of `C', i.e. sources fitted with a single Gaussian 
component, but which lie in the same island as other source(s). There are about 
$18$k compact `M' sources, $10$k of which are isolated. Such sources may be 
unambiguous single sources with substructure (e.g. the two lobes of a radio 
galaxy) or may be two or more nearby distinct sources that have been grouped as 
a single source by \PyBDSF/, i.e. blended sources. An additional complication is 
that calibration errors and dynamic range limitations lead to shape distortions, 
resulting in multiple Gaussian components being fitted by \PyBDSF/ to a single 
source. Moreover, true extended radio sources are not necessarily Gaussian in 
shape or even composed of the sum of Gaussian shapes. This is a choice of 
representation imposed by our source detection algorithm. These factors, and 
intrinsic asymmetries in the sources (e.g.\ head-tail sources), mean that even 
in the case of single sources, the flux-weighted source positions provided by 
\PyBDSF/ may not coincide with the optical host galaxy positions, making the LR 
values unreliable. Nevertheless, combining the information in the LR matches to 
both the overall source and to the individual Gaussian components provides a 
means to diagnose specific cases and either allow an LR result to be obtained 
for a source or to identify cases for further visual inspection and deblending.

\begin{figure*}
 \centering
 \includegraphics[width=0.85\textwidth]{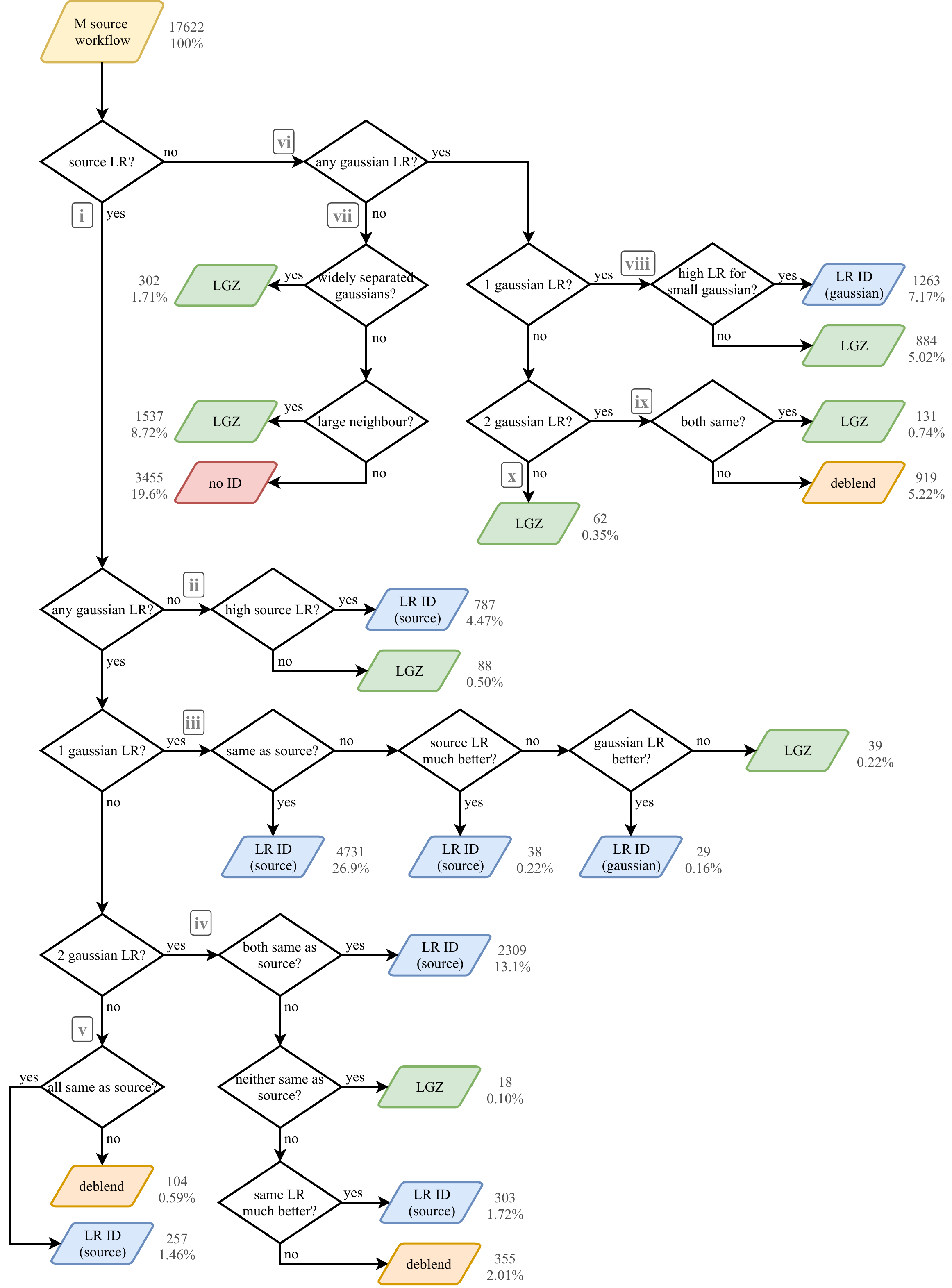}
 \caption{High level summary of the decision tree used to process all compact 
`M' sources (i.e. \PyBDSF/ sources fitted with multiple Gaussians) in the  
\PyBDSF/ catalogue. Following this workflow a decision is made for each source 
whether to: (i) make the optical/IR identification, or lack thereof, through the 
LR method (blue and red outcomes respectively) for either the \PyBDSF/ source or 
one of the Gaussian components; (ii) process the source in LGZ  (green 
outcomes); or (iii) process further in a separate deblending workflow (orange 
outcomes, see Section\ \ref{sec:deblend}).  The key parameters are defined in 
Table~\ref{tab:cat:flowkeyM} and full details of the decisions are given in 
Section\ \ref{sec:cat:flow:m}, with reference to the branch labels i--x. The 
numbers reflect the number of \PyBDSF/ sources in each final bin and the 
percentage is relative to the total number of  compact `M' sources in the 
\PyBDSF/ catalogue. }
 \label{fig:cat:flowchartM}
\end{figure*}

These compact `M' sources may be isolated or not, but were treated in a separate 
`M source' workflow, in which we also considered any LR matches to the 
individual Gaussian components of each source. A schematic overview of this 
decision tree is given in Fig.~\ref{fig:cat:flowchartM}, and key parameters are 
defined in Table\ \ref{tab:cat:flowkeyM} and described in detail in the 
following subsections. The only difference between isolated and non-isolated `M' 
sources is that non-isolated sources were subjected to additional visual sorting 
before inclusion in the LGZ sample; for clarity this is not shown explicitly in 
Fig.~\ref{fig:cat:flowchartM}, but each decision that ends in `LGZ' for the 
non-isolated sources can be taken to mean `visually confirmed for LGZ' otherwise 
the alternate decision was followed, while for the isolated sources they were 
directly added to the LGZ sample.   The final decisions were to accept the 
source LR match, accept one of the Gaussian LR matches, include the source in 
the LGZ sample (where one of the possible outcomes is blended; see Section 
\ref{sec:cat:lgz}), or pass the source directly to a separate deblending 
workflow (see Section~\ref{sec:deblend}). 

\begin{table*}[htp]
 \centering
 \caption{Definition of the parameters used in the decision tree for `M' sources 
 (i.e. \PyBDSF/ sources fitted with multiple Gaussians) in 
Fig.~\ref{fig:cat:flowchartM}.  See Section\ \ref{sec:cat:flow:m} for details.}
 \small
 \label{tab:cat:flowkeyM}
\begin{tabular}{ll}
\hline
\hline
Parameter & definition \\
\hline
Source LR & $LR_{\mathrm{source}} > 0.639$ \\
Any Gaussian LR & at least one Gaussian component has $LR_{\mathrm{gauss}} > 
0.639$\\
High source LR &  $LR_{\mathrm{source}} > 6.39$ \\
$N$ Gaussian LR & $N$ Gaussian components have $LR_{\mathrm{gauss}} > 0.639$\\
Same as source & the ID(s) for the Gaussian component(s) are identical to the  
ID for the source\\
Source LR much better &  $LR_{\rm source} > 10$ \& $LR_{\rm gauss} < 10$ \& 
$LR_{\rm source} > 10 LR_{\rm gauss}$ \\
Gaussian LR better  & $LR_{\rm gauss} > 10$ \& $LR_{\rm source} < 10$ \& 
$LR_{\rm gauss} > 10 LR_{\rm source}$\\
Same LR much better & $LR_{\rm source} > 100$ \& $LR_{\rm gauss (same)} > 5 
LR_{\rm gauss (different)}$ \\
Widely separated Gaussians & maximum separation between Gaussian components of 
the \PyBDSF/ source larger than $15${\arcsec} \\
Large neighbour & large ($> 10${\arcsec}) neighbour within $100${\arcsec}  \\
High LR for small Gaussians  & $LR_{\rm gauss} >6.39$ \& the Gaussian size is 
$<10${\arcsec} \\
\hline
\end{tabular}
\end{table*}

\subsubsection{Sources with a LR identification}

We first considered whether the source has a LR match above the threshold 
(branch i in Fig.~\ref{fig:cat:flowchartM}), then whether at least one of the  
Gaussian components has an LR match above the threshold, and subsequently tried 
to resolve any ambiguities in the optical matches to the source and Gaussian 
components. If none of the Gaussian components have a good LR match (branch ii), 
then the source match was accepted provided the LR exceeded a higher threshold; a 
threshold ten times normal ($LR > 10 L_{\rm thr} = 6.39$) was used because `M' sources often have larger uncertainties on their 
source positions, which can lead to lower LR misidentifications, especially for 
sources lying in over-dense environments. Otherwise the source was included in 
the LGZ sample for closer inspection. A high threshold source LR match with no 
good Gaussian LR matches generally occurs when a slightly resolved double radio 
source is composed of two (or more) Gaussian components which are correctly 
grouped by \PyBDSF/ as a source whose position corresponds to the optical ID.

When only one Gaussian component has an LR identification (branch iii), the 
majority of the time it is the same optical/IR source as the source match and 
the identification is unambiguous. In the remaining few cases, where the single 
Gaussian component LR match is different to the source match, we evaluated 
whether one is significantly better than the other. The source match was 
accepted only if the source LR exceeds a higher threshold and exceeds ten times 
that of the Gaussian component: $LR_{\rm source} > 10$ \& $LR_{\rm gauss} < 10$ 
\& $LR_{\rm source} > 10 LR_{\rm gauss}$. Likewise, the Gaussian component LR 
match was preferred to that of the overall source under the reverse conditions: 
$LR_{\rm gauss} > 10$ \& $LR_{\rm source} < 10$ \& $LR_{\rm gauss} > 10 LR_{\rm 
source}$. These ranges were chosen empirically based on visual inspection of 
images of subsamples of sources. The remaining in-between cases, where neither 
the source nor the Gaussian component LR match can be deemed to be reliably 
better by statistical methods only, were evaluated in LGZ.  

The situation is more complex if more than one Gaussian component has an 
acceptable LR match. In roughly three-quarters of the cases in which  two 
Gaussian components have LR matches  (branch iv),  both Gaussian components 
match the same optical source as the radio source LR match, so the 
identification could be unambiguously accepted. Another quarter fall into the 
category where one Gaussian component match is the same as the source match. 
Some of these were deemed to have a very good source match ($LR_{\rm source} > 
100$ \& $LR_{\rm gauss (same)} > 5 LR_{\rm gauss (different)}$) and so the 
source match was accepted. The rest were processed in the deblending workflow. 
For a small number of sources, the two Gaussian component matches and source 
match are all different. These sources were processed in LGZ. 

Finally, only a small number of sources have three or more Gaussian components 
with LR matches (branch v), and in this case about three quarters all 
unambiguously match  the same optical source as the source LR match, which was 
then accepted, while the remainder were processed via the deblending workflow.

\subsubsection{Sources without a LR identification}
The second major branch of this decision tree considers the case where there is 
no good LR match to the overall source (branch vi). If these are isolated 
sources, then they may simply have no counterpart above the sensitivity limits 
of the Pan-STARRS/WISE data. But equally these may be asymmetric sources where 
the flux-weighted source position does not accurately coincide with the optical 
counterpart location or these sources may have clear substructure where one 
of the Gaussian components may coincide with the optical counterpart. 
Alternatively, these may be blended sources. To assess these possibilities, we 
again considered whether, and how many, Gaussian components have acceptable LR 
matches. In the case where no Gaussian components have LR matches (branch vii), 
it is very likely that the source of interest has no optical/IR identification. 
However, we also consider cases where the source may be complex or a component 
of a larger structure. Thus, sources whose Gaussian components are widely 
separated (maximum separation larger than $15${\arcsec}) or sources that have an 
extended ($>10${\arcsec}) neighbouring radio source within $100${\arcsec} were 
included in the LGZ sample. This is the only step where the decision differs 
significantly for the isolated and non-isolated `M' sources, where, by 
definition, a much higher fraction of non-isolated sources would be included in 
the LGZ sample. A visual sorting step was done on the non-isolated sources 
selected for LGZ, to avoid adding too many trivial sources with no optical/IR 
identification (again for clarity this is not shown explicitly in 
Fig.~\ref{fig:cat:flowchartM}). 

If only one Gaussian component has an acceptable LR match (branch viii) it was 
taken as the source match, provided it was deemed a good match ($LR_{\rm gauss} 
>10 LR_{\rm thresh}$ and the Gaussian size is $<10${\arcsec}); these limits were 
again determined by visual inspection of subsamples of sources, in that a lower 
LR threshold or larger size threshold produced too many wrong matches while 
everything satisfying these criteria appeared to be genuine. Otherwise, the 
source was included in the LGZ sample. 

Where there were two Gaussian components with acceptable LR matches (branch ix), 
if these matched to the same optical galaxy the source was handled in LGZ; this 
is because the lack of a good source LR match on this branch, combined with the 
two acceptable Gaussian LR matches, while likely to be the correct match, 
suggests some complex structure may be present. Otherwise if there were two 
separate optical galaxies, there is a strong possibility that the components 
were mistakenly grouped as a single source by \PyBDSF/ and so the \PyBDSF/ source was 
examined in the deblending workflow. 
Finally, the few sources with three or more Gaussian components with good LR 
matches were processed in LGZ (branch x).

\section{Final catalogue}
\label{sec:cat:final}

A final catalogue of LoTSS radio sources cross-matched to 
Pan-STARRS/\textit{WISE} was produced by combining the identifications (and 
associations) from all the identification methods, including the LR method, LGZ, 
the deblending workflow, and the large galaxies.  In the following, associated 
sources 
refer to those where separate \PyBDSF/ sources have been associated and 
combined 
into single new sources either based on the LGZ output or matches with large 
optical galaxies (see Section\ \ref{sec:cat:flow:bright}). The
individual \PyBDSF/ sources that make up
associated sources were removed from the catalogue and replaced with the 
associated sources. All artefacts,  identified at various stages of the 
decision tree and LGZ, were removed from the catalogue \citepalias[and also 
flagged as such in the catalogue of][]{Shimwell_2018}. Sources that were 
identified as blends and processed 
in the deblending workflow were  also removed and replaced by sources made 
up of one or more Gaussian components (see Section\ \ref{sec:deblend}); they 
therefore have properties appropriate for associated sources in the catalogue. 
For all associated sources, we generated the LoTSS source properties and 
populated the appropriate final catalalogue columns
(e.g. total flux density, size, radio position, and radio source name) by
combining the \PyBDSF/ properties of their constituent components (or
Gaussian components in the case of blends) as described in Section\ 
\ref{sec:assoc}. 

The LoTSS-DR1 value-added catalogue lists the radio properties, identification 
methods, and optical properties where available. The columns in the catalogue 
describing the LoTSS properties are as follows \citepalias[for more details 
see][]{Shimwell_2018}:
\begin{itemize}
\item The IAU source identification (`Source\_Name') based on the position of 
each source.
\item LoTSS position and errors (`RA', `E\_RA', `DEC', and `E\_DEC'). In the 
case of associated sources, this is the flux-weighted mean of the component 
values.
\item LoTSS peak and total flux densities and associated errors (`Peak\_flux', 
`E\_Peak\_flux', `Total\_flux', and `E\_Total\_flux'). In the case of associated 
sources this is the maximum of the peak flux densities and sum of the total flux 
densities of the components.
\item LoTSS shape (`Maj', `E\_Maj', `Min', `E\_Min', `PA', `E\_PA') and  
deconvolved shape (`DC\_Maj', `E\_DC\_Maj', `DC\_Min', `E\_DC\_Min', `DC\_PA', 
`E\_DC\_PA'). Deconvolved values are zero for unresolved sources. All these 
values are blank for associated sources whose shapes are described in different 
columns outlined below.
\item Local rms noise in the LoTSS map (`Isl\_rms'). In the case of associated 
sources this is the mean value of the components.
\item Multiple Gaussian code (`S\_Code') is `M' in the case where the source 
consists of multiple Gaussian components or associated sources, `S' where it 
consists of a single Gaussian, and `C' in the case where the source lies within 
the same island as another source. These codes are updated for the sources that 
are associated or deblended. 
\item Name of the LoTSS mosaic in which the source can be found  (`Mosaic\_ID').
\item The ratio of the number of LoTSS pointings in which the source  is in the 
\textsc{Clean} mask to the number of pointings which are mosaicked at the 
position of the source  (`Masked\_Fraction').
\end{itemize}
The associated sources have values for the following additional columns for 
their LoTSS properties determined as described in Section~\ref{sec:assoc} (these 
are blank for non-associated sources):
\begin{itemize}
\item Shape measurements for associated sources (`LGZ\_Size', `LGZ\_Width', 
`LGZ\_PA').
\item The number of \PyBDSF/ sources in the association `LGZ\_Assoc'.
\item A quality flag for the association (`LGZ\_Assoc\_Qual'). For LGZ
  this is the fraction of all views of this source region for which
  the listed association was chosen as the best associated set. Only
  sets with LGZ\_Assoc\_Qual$>2/3$, and, of those, only the
    largest set for each LGZ input source are included in the final
  catalogue, with a small number of overlapping association sets
  resolved visually (see Section~\ref{sec:cat:lgz:out}). This flag is
  set to 1 for the sources automatically associated based on a bright
  galaxy match or in the deblending workflow.
\end{itemize}
Information pertaining to the optical/IR identification is given by the 
following:
\begin{itemize}
\item A flag indicating the origin of the optical/IR identification or 
non-identification (`ID\_flag'). The description of these flags can be found in 
Table\ \ref{tab:id_flags}. For ID\_flag=0, no attempt is made at an 
identification, while for the other values, the ID\_flag indicates only which 
method was used to attempt an identification and not whether an ID is made. For 
example, a source with ID\_flag = 1 may have an optical/IR identification above 
the LR threshold or it may have no acceptable LR identification.
\item Name (`ID\_name') and position (`ID\_ra' and `ID\_dec') of the optical/IR 
identification, when present (sources with no identification can be recognised 
because they have no ID\_name, ID\_ra and ID\_dec values). The recorded values 
are the Pan-STARRS object name and position or the AllWISE source name and 
position in the case of no Pan-STARRS detection. A small number (1078) of 
sources with a match to a bright galaxy (either through the decision tree or LGZ 
`host broken up') have an ID\_name from 2MASX or SDSS, while the position is 
taken from the nearest match for that 2MASX or SDSS source in the merged 
Pan-STARRS/AllWISE catalogue, with the caveat that the PanSTARRS and AllWISE 
photometry is likely to be wrong for these large sources.
\item The LR for sources where the identification is made through this maximum 
likelihood method (`ML\_LR').
\item A quality flag for LGZ identifications (`LGZ\_ID\_Qual'). This is set to 
the fraction of all LGZ views of this source region for which the catalogued ID 
was selected. Only IDs with LGZ\_ID\_Qual$>2/3$, and only the highest quality ID 
for each source, were included in the catalogue.
\item For deblended sources, the name of the \PyBDSF/ multiple Gaussian 
component source from which each source was deblended (`Deblended\_from'). This 
is blank for all other sources.
\end{itemize}
For the sources that have optical/IR identifications, we include the Pan-STARRS 
and AllWISE photometry:
\begin{itemize}
 \item The name of the source in the AllWISE catalogue, `AllWISE'.
 \item The Pan-STARRs object ID, `objID'.
 \item Pan-STARRS forced aperture fluxes, magnitudes, and errors in the 
Pan-STARRS $grizy$ bands  (`<band>FApFlux', `<band>FApFluxErr', `<band>FApMag', 
and `<band>FApMagErr').
 \item Pan-STARRS Kron fluxes and errors in the Pan-STARRS $grizy$ bands  
(`<band>FKronFlux' and `<band>FKronFluxErr').
 \item AllWISE profile fitted  fluxes, magnitudes, and errors in the 
\textit{WISE} W1, W2, W3, and W4 bands (`<band>Flux', `<band>FluxErr', 
`<band>Mag', and `<band>MagErr'). Sources with zero `Flux' values in a 
particular band were not detected in that band, and they have a $1\sigma$ upper 
limit given in the `FluxErr' column.
\end{itemize}
Additional columns pertaining to the photometric redshifts and rest-frame 
colours are described by \citetalias{Duncan_2018}.

\begin{table*}
\centering
\caption{Descriptions of the ID\_flag keyword in the final catalogues used to 
indicate the origin of the possible association and optical/IR identification, 
or lack thereof.}
\small
\label{tab:id_flags}
\begin{tabular}{lp{14cm}}
\hline
\hline
ID\_flag & description  \\
\hline
0 &  no identification is possible -- in cases of extended diffuse emission\\
1 &  the identification (or lack thereof) is made through the LR method \\
2 &  the identification, and possible association, is made based on a match to a 
bright optical/IR galaxy \\
22 &  the identification is made based on a match to a bright optical/IR galaxy  
after classification, and possible association, in LGZ as `host broken up' \\
31 & the  possible association and identification (or lack thereof) is made 
through LGZ \\
32 & the  possible association and identification (or lack thereof) is made  
after further processing when classified as `too zoomed in' in LGZ (on branch B 
of the decision tree) \\
41 & the source and identification (or lack thereof) comes from the deblending 
workflow based on the `M' source decision tree \\
42 & the source and identification (or lack thereof) comes from the deblending 
workflow after classification in LGZ as `blend' \\
\hline
\end{tabular}
\end{table*}

We also retain a component catalogue of the sources in the \PyBDSF/ catalogue  
associated as components in the final LoTSS-DR1 value-added catalogue. Each 
entry in the component catalogue has an identifier `Component\_Name' based on 
the component position in the \PyBDSF/ catalogue and a `Source\_Name', which 
corresponds to that in the value-added catalogue. The component catalogue 
includes a column, `Ng', that gives the number of Gaussian components in each 
source. It also includes the additional \textsc{Clean} mask columns, 
`Number\_Masked', and `Number\_Pointings', giving the number of LoTSS pointings 
in which the source  is in the \textsc{Clean} mask and the number of pointings 
which are mosaicked at the position of the source 
\citepalias[see][]{Shimwell_2018}. Each deblended source also appears as a 
component in the components catalogue;  for these sources, we include the 
column, `Deblended\_from', which gives the name of the \PyBDSF/ multiple 
Gaussian component source from which each source was deblended.

The catalogues presented in this paper are now publicly available\footnote{The 
LoTSS-DR1 images and catalogues, including the value-added catalogue presented 
here, can be found at \url{https://lofar-surveys.org}.}.  The final catalogue 
contains 318,520 radio sources, of which 231,716 (73\%) have optical/IR 
identifications.  Table~\ref{tab:id_fractions_type} shows the total number  of 
sources, as well as the number and fraction of sources with an identification, 
for the different identification methods.  The majority of the identifications 
come from the LR method with an overall identification rate of $74$\%. The 
overall identification rate for the LGZ method is $60$\%.  Sources identified on 
the basis of a bright optical galaxy have $100$\% identifications by 
construction.  The deblending route has a high identification rate as sources 
are generally only selected for deblending when there are clear optical/IR 
identifications for several of the components.

\begin{table}
\centering
\caption{Total number of sources and the number with identifications for each 
method of identification.}
\small
\label{tab:id_fractions_type}
\begin{tabular}{lrrl}
\hline
\hline
& Number  & Number & ID  \\
&         & with ID  & fraction \\
\hline
All Sources    & 318,520  &  231,716 & 0.73 \\
LR             & 299,730  &  221,269 & 0.74 \\
LGZ            & 11,989   &  7,144   & 0.60 \\ 
Deblending     & 2,435    &  2,338   & 0.96 \\
Bright galaxy  & 965      &  965     & 1.00 \\
No ID possible & 3,401    &  0       & 0.00 \\
\hline
\end{tabular}
\end{table}

\begin{figure}
\centering
  \includegraphics[width=0.5\textwidth]{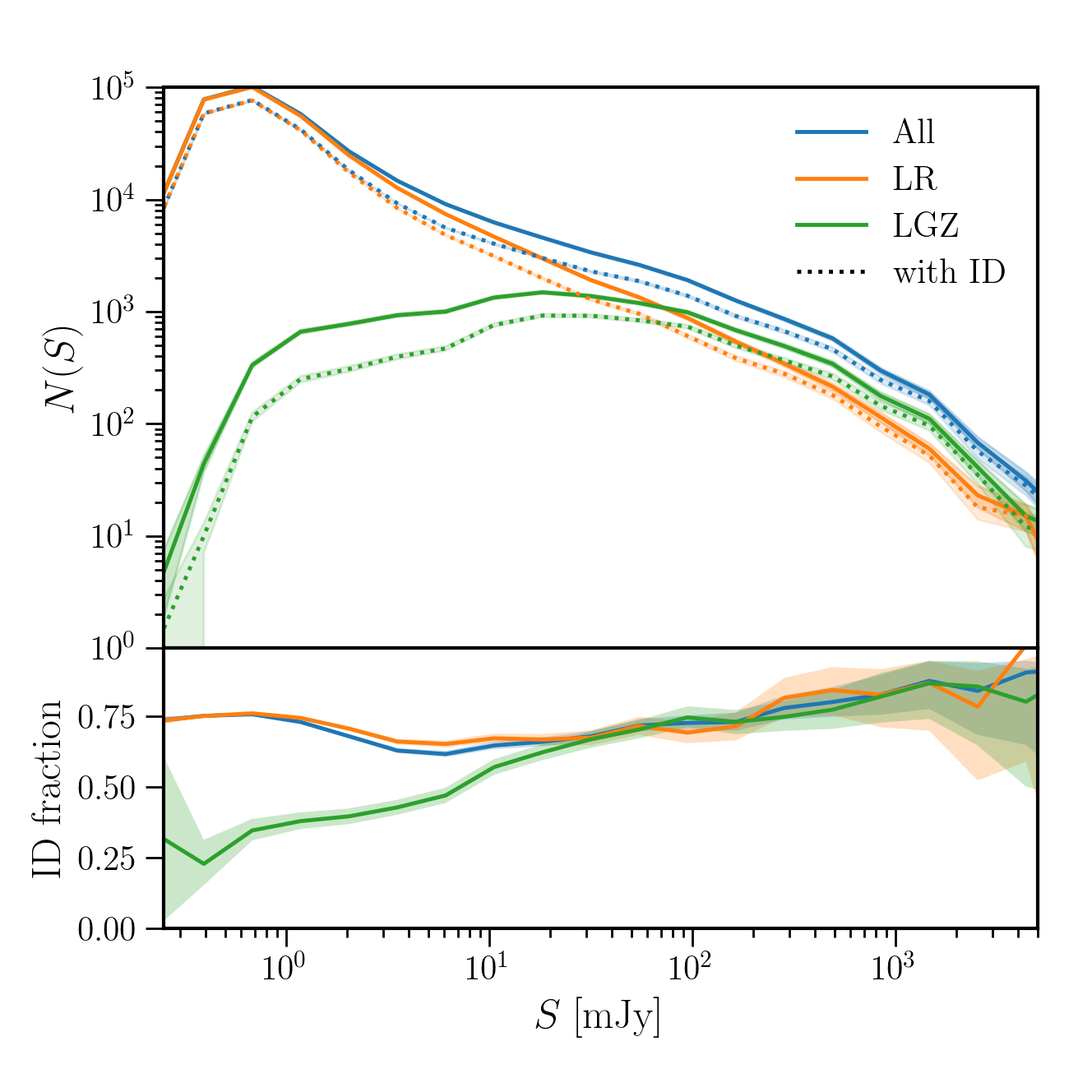}
  \caption{Total number of sources (solid lines) and number of sources with 
identifications (dotted lines) as a function of 150-MHz flux density, in bins of 
$0.23$\,dex, for all sources (blue) and via the two major methods: LGZ  (green) 
and LR (orange). The respective fractions of identifications (i.e. the ratio of 
the number of sources with identifications in each category to the number of 
sources in each category) as a function of flux density are shown in the bottom 
panel. Filled regions show the errors that are estimated using Monte Carlo 
simulations drawn from Poissonian distributions. }
  \label{fig:id_fraction}
\end{figure}

The number of sources and identification fractions for the LR and LGZ methods 
are shown as a function of flux density in Fig.~\ref{fig:id_fraction}. The 
identification fraction here is the ratio of the number of sources with 
identifications to the number of sources in that category, and therefore shows 
the variation in identification rate as a function of flux density for each 
method. Errors on the numbers and fractions, within each flux density bin, were 
estimated using Monte Carlo simulations drawn from Poissonian distributions; for 
large numbers this converges to the Gaussian distribution. The LGZ 
identification fraction drops from 75\% for sources with flux densities above 
100\,mJy down to below 25\% at the lowest flux densities. The decrease in LGZ 
identification at low flux densities can be explained by the fact that by 
construction the sources selected for LGZ processing are resolved and those at 
lower flux densities are more likely to be AGN at high redshifts whose host 
galaxies fall below the optical/IR flux limits of Pan-STARRS/AllWISE.

\begin{figure}
\centering
  \includegraphics[width=0.5\textwidth]{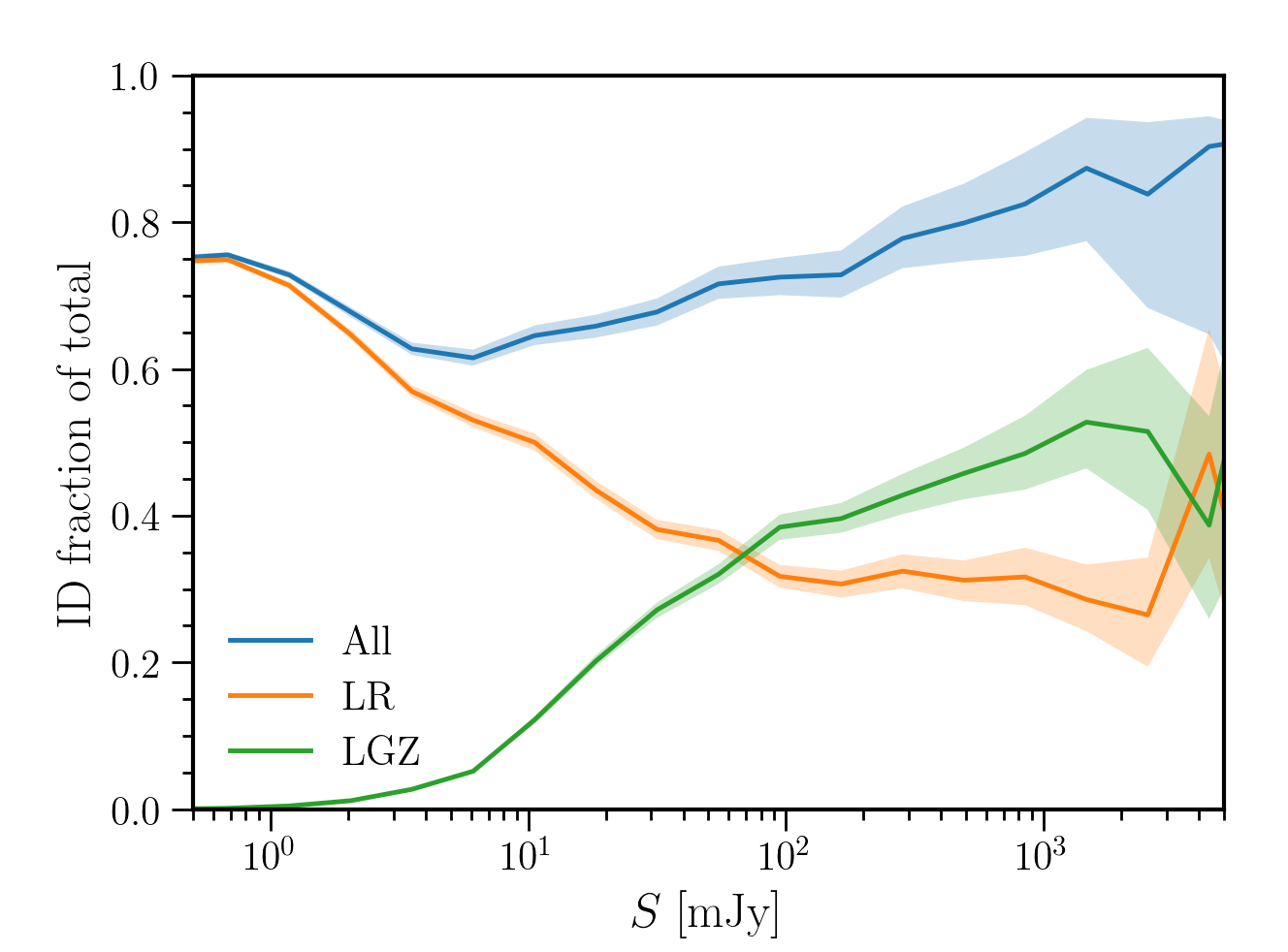}
  \caption{Contribution to the overall identification fraction (i.e. the ratio 
of the number of sources with identifications within each category to the total 
number of sources) for sources at a given 150 MHz  flux density,  in bins of 
$0.23$\,dex,  for all sources (blue) and via the two major methods: LGZ  (green) 
and LR (orange). Filled regions show the errors that are estimated using Monte 
Carlo simulations drawn from Poissonian distributions. }
  \label{fig:id_fraction2}
\end{figure}

Fig.~\ref{fig:id_fraction2} shows the relative contribution by the two main 
identification methods to the overall identification fraction for all sources as 
a function of 150-MHz flux density, i.e. the ratio of the number of sources with 
identifications within each category to the total number of sources. This shows 
the contribution of each identification method to the total identification rate 
as a function of flux density, highlighting the fact that the majority of the 
optical/IR identifications for radio sources above a few tens of mJy come from  
LGZ, while those for fainter sources come from the LR method. Interestingly, the 
overall identification fraction drops with decreasing flux density down to 
$\approx 5$\,mJy, but then rises again at lower flux densities. These properties 
can be easily understood by considering the different radio source populations 
at different flux densities. At the brightest flux densities, the radio source 
counts are dominated by powerful radio-loud AGN, which often have extended 
complex radio structures requiring LGZ analysis. As the flux density decreases, 
the average redshift of these radio-loud AGN increases, leading to more of the 
optical counterparts falling below the magnitude limit of the Pan-STARRS and 
\textit{WISE} catalogues and a decreasing overall ID fraction. At flux densities 
below a few mJy, however, the dominant contribution to the overall radio 
population switches: star-forming galaxies begin to dominate the radio source 
counts \citep[e.g.][]{wilman_2008,padovani_2016,Williams_2016}. These are mostly 
at lower redshift, with consequently brighter counterparts, and are largely 
single radio components matching the counterpart position; this leads to an 
increasing proportion of the overall population for which IDs are found with 
most of these IDs coming from LR\textbf{}s.

\section{Summary and future prospects}
\label{sec:summary}

In this paper we have presented a catalogue of optical/IR identifications for 
radio sources in the first LoTSS data release  presented by 
\citet[][DR1-I]{Shimwell_2018}. We have used a statistical  colour- and 
magnitude-dependent LR method for the cross-matching of the majority of sources, 
complemented by LOFAR Galaxy Zoo (LGZ), a Zooniverse-based visual association 
and identification project, for sources with complex structure. The LGZ method, 
while time consuming, is well suited both for characterising large radio sources 
as well as identifying their optical/IR counterparts. The LR method cannot be 
used for such sources, but is an efficient way to identify the likely hosts of 
the majority of the LoTSS radio sources. We have therefore made use of a 
decision tree, based on the radio source properties and LRs, to select complex 
sources for visual classification with LGZ. This approach, of reserving the 
complex sources for visual classification while using statistical methods for 
the majority of sources, may be useful for future wide-area radio surveys.

The final radio source catalogue contains 318,520 entries, of which 231,716 
($73$\%) have optical/IR identifications from Pan-STARRS and/or \textit{WISE} 
(or in a few cases, 2MASX or SDSS). Most of the identifications at the brighter 
radio flux densities come from LGZ, while those at lower flux densities come 
from LR.  In both cases, the identification rates depend on the quality and 
depth of the multiwavelength data and the underlying radio source population.

At just over 400 square degrees, LoTSS-DR1 covers only about 2\% of the total 
sky area expected to be covered by LoTSS. Additionally, the LOFAR surveys will 
include deeper tiers covering smaller areas. Fortunately, the source population 
at fainter flux densities mostly comprises star-forming galaxies or faint 
unresolved AGN, which are well suited for  cross-matching using the LR method 
(although a small number require deblending). The large increase in source 
numbers, in particular of the complex sources, within the full LoTSS coverage 
will require a different approach. This may involve an expansion of our LGZ 
Zooniverse project to the public, similar to  `Radio Galaxy Zoo' 
\citep{Banfield_2015}, which is using citizen scientists to cross-match over 
170,000 radio sources. Work has been done on automated algorithms that can 
perform the cross-matching of complex radio sources 
\citep[e.g.][]{Proctor_2006,vanVelzen_2015,Fan_2015}, but these have mostly used 
simple pattern recognition algorithms that will only identify the simplest, most 
common, cases (e.g. well-defined double or triple sources). More recent work 
involves machine learning techniques such as self-organising maps or Kohonen 
maps  \citep[e.g. parallelised rotation/flipping INvariant Kohonen maps, or 
PINK;][]{Polsterer_2015} to construct prototypes of radio galaxy morphologies, 
which are being applied to the LoTSS data  \citep{Mostert_2017}. 
\cite{Aniyan_2017} have used convolutional neural networks  to classify radio 
galaxy images into \cite{FR_1974} Type 1 or 2 (FRI/FRII)  classes.  Similarly, 
\cite{Lukic_2018} have classified radio galaxy morphologies in  distinct 
classes, optimising the convolutional neural network parameters to produce four 
classes consisting of compact, single-, double-, and multiple-component extended 
sources. While  many of these efforts are still focussed on the morphological 
classification of the radio structures and not on the optical/IR identification, 
they do allow a means to identify similar cases where the identification can be 
made relatively easily with automated algorithms, and outliers which may require 
human intervention to identify any counterparts.

The value of these identifications are further enhanced by estimates of the 
distances (via redshifts), from which one can calculate instrinsic properties 
such as luminosities and physical sizes. Photometric redshift and rest-frame 
colour estimates for all radio sources with identified optical counterparts 
presented in this paper are provided by \citet[][DR1-III]{Duncan_2018}. In the 
future, spectroscopic surveys such as WEAVE-LOFAR \citep{Smith_2016} will 
provide precise redshift estimates and robust source classification for large 
numbers of the LoTSS source population.

\section*{Acknowledgements}

% lofar credits
This paper is based on data obtained with the International LOFAR Telescope 
(ILT) under project codes LC2\_038 and LC3\_008. LOFAR 
(\citealt{vanHaarlem_2013}) is the LOw Frequency ARray designed and constructed 
by ASTRON. It has observing, data processing, and data storage facilities in 
several countries, which are owned by various parties (each with their own 
funding sources) and are collectively operated by the ILT foundation under a 
joint scientific policy. The ILT resources have benefited from the following 
recent major funding sources: CNRS-INSU, Observatoire de Paris and 
Universit\'{e} d'Orl\'{e}ans, France; BMBF, MIWF-NRW, MPG, Germany; Science 
Foundation Ireland (SFI), Department of Business, Enterprise and Innovation 
(DBEI), Ireland; NWO, The Netherlands; The Science and Technology Facilities 
Council, UK[7].

Part of this work was carried out on the Dutch national e-infrastructure with 
the support of SURF Cooperative through grant e-infra 160022 and we gratefully 
acknowledge support by N. Danezi (SURFsara) and C. Schrijvers (SURFsara).

This research has made use of the University of Hertfordshire
high-performance computing facility (\url{http://uhhpc.herts.ac.uk/}) and the 
LOFAR-UK computing facility located at the University of Hertfordshire and 
supported by
STFC [ST/P000096/1]. 

This research made use of \textsc{astropy}, a community-developed core {\sc 
Python} package for astronomy \citep{2013A&A...558A..33A} hosted at 
{\url{http://www.astropy.org/}}, of \textsc{APLpy}  \citep{Robitaille_2012}, an 
open-source astronomical plotting package for {\sc Python} hosted at 
{\url{http://aplpy.github.com/}}, and of \textsc{topcat} 
\citep{2005ASPC..347...29T}.

% panstarrs credits
The Pan-STARRS1 Surveys (PS1) and the PS1 public science archive have been made 
possible through contributions by the Institute for Astronomy, the University of 
Hawaii, the Pan-STARRS Project Office, the Max-Planck Society and its 
participating institutes, the Max Planck Institute for Astronomy, Heidelberg and 
the Max Planck Institute for Extraterrestrial Physics, Garching, the Johns 
Hopkins University, Durham University, the University of Edinburgh, the Queen's 
University Belfast, the Harvard-Smithsonian Center for Astrophysics, the Las 
Cumbres Observatory Global Telescope Network Incorporated, the National Central 
University of Taiwan, the Space Telescope Science Institute, the National 
Aeronautics and Space Administration under Grant No. NNX08AR22G issued through 
the Planetary Science Division of the NASA Science Mission Directorate, the 
National Science Foundation Grant No. AST-1238877, the University of Maryland, 
Eotvos Lorand University (ELTE), the Los Alamos National Laboratory, and the 
Gordon and Betty Moore Foundation.

% wise credits
AllWISE makes use of data from {\it WISE}, which is a joint project of the 
University of California, Los Angeles, and the Jet Propulsion 
Laboratory/California Institute of Technology, and NEOWISE, which is a project 
of the Jet Propulsion Laboratory/California Institute of Technology. {\it WISE} 
and NEOWISE are funded by the National Aeronautics and Space Administration.

% Zooniverse
This publication uses data generated via the \url{Zooniverse.org} platform, 
development of which is funded by generous support, including a Global Impact 
Award from Google, and by a grant from the Alfred P. Sloan Foundation.

% author funding
WLW and MJH acknowledge support from the UK Science and Technology Facilities 
Council (STFC) under grant ST/M001008/1.
PNB and JS are grateful for support from the UK STFC via grant ST/M001229/1.
JHC and BM acknowledge support from the UK STFC under grants ST/M001326/1 and 
ST/R00109X/1.
RKC, CLH, and RK acknowledge support from STFC studentships.
VHM thanks the University of Hertfordshire for a research studentship 
[ST/N504105/1]. 
LA acknowledges support from the STFC through a ScotDIST Intensive Data Science 
Scholarship.
GJW gratefully acknowledges support from the Leverhulme Trust.
LKM acknowledges the support of the Oxford Hinzte Centre for Astrophysical 
Surveys, which is funded through generous support from the Hintze Family 
Charitable Foundation. This publication arises from research partly funded by 
the John Fell Oxford University Press (OUP) Research Fund. 
GGU acknowledges support from the CSIRO OCE Postdoctoral Fellowship.
The LOFAR group at Leiden acknowledges support from the ERC Advanced 
Investigator programme NewClusters 321271. 
RJvW further acknowledges support from the VIDI research programme with project 
number 639.042.729, which is financed by the Netherlands Organisation for 
Scientific Research (NWO).
FdG is supported by the VENI research programme with project number 1808, which 
is financed by the NWO.
APM would like to acknowledge support from the NWO/DOME/IBM programme ``Big Bang 
Big Data: Innovating ICT as a Driver For Astronomy'', project \#628.002.001.
AG acknowledges full support from the Polish National Science Centre (NCN) 
through the grant 2012/04/A/ST9/00083.
MKB acknowledges support from the Polish National Science Centre under grant no. 
2017/26/E/ST9/00216.
IP acknowledges support from INAF under PRIN SKA/CTA `FORECaST'.

\bibliographystyle{aa}
\bibliography{LoTSS-HETDEX-IDs}

\tiny
\noindent\rule{\linewidth}{0.4pt}
$^{1}$Centre for Astrophysics Research, School of Physics, Astronomy and 
Mathematics, University of Hertfordshire, College Lane, Hatfield AL10 9AB, UK\\
$^{2}$SUPA, Institute for Astronomy, Royal Observatory, Blackford Hill, 
Edinburgh, EH9 3HJ, UK\\
$^{3}$School of Physical Sciences, The Open University, Walton Hall, Milton 
Keynes, MK7 6AA, UK\\
$^{4}$Leiden Observatory, Leiden University, PO Box 9513, NL-2300 RA Leiden, The 
Netherlands \\
$^{5}$ASTRON, the Netherlands Institute for Radio Astronomy, Postbus 2, 7990 AA, 
Dwingeloo, The Netherlands\\
% group 2
$^{6}$CSIRO Astronomy and Space Science, PO Box 1130, Bentley WA 6102, Australia 
\\
$^{7}$Astronomical Observatory, Jagiellonian University, ul. Orla 171, 30-244 
Krak\'ow, Poland \\
$^{8}$Astrophysics, University of Oxford, Denys Wilkinson Building, Keble Road, 
Oxford, OX1 3RH, UK \\
$^{9}$Jodrell Bank Centre for Astrophysics, School of Physics and Astronomy, 
University of Manchester, Manchester M13 9PL, UK \\
$^{10}$Toru\'n Centre for Astronomy, Faculty of Physics, Astronomy and 
Informatics, NCU, Grudziacka 5, 87-100 Toru\'n, Poland \\
$^{11}$INAF -- Istituto di Radioastronomia, via Gobetti 101, 40129 Bologna, 
Italy \\
$^{12}$Anton Pannekoek Institute for Astronomy, University of Amsterdam, Postbus 
94249, 1090 GE Amsterdam, The Netherlands\\
$^{13}$GEPI, Observatoire de Paris, CNRS, Universite Paris Diderot, 5 place 
Jules Janssen, 92190 Meudon, France\\
$^{14}$Department of Physics \& Electronics, Rhodes University, PO Box 94, 
Grahamstown, 6140, South Africa\\
$^{15}$Space Science \& Technology Department, The Rutherford Appleton 
Laboratory, Chilton, Didcot, Oxfordshire OX11 0NL, UK\\
$^{16}$University of Hamburg, Hamburger Sternwarte, Gojenbergsweg 112, 21029 
Hamburg, Germany \\
$^{17}$Th\"{u}ringer Landessternwarte (TLS), Sternwarte 5, D-07778 Tautenburg, 
Germany \\
$^{18}$Kapteyn Astronomical Institute, PO Box 800, NL-9700 AV Groningen, the 
Netherlands \\
$^{19}$Fakult\"at f\"ur Physik, Universit\"at Bielefeld, Postfach 100131, 33501 
Bielefeld, Germany \\

\label{lastpage}
\end{document}